\documentclass[11pt, a4paper]{article}
\usepackage{jheppub}
\pdfoutput=1
\usepackage{bm, mathrsfs, physics, slashed, tensor, cancel, mathtools}
\usepackage{tikz, caption, subcaption, float}

\title{Holographic mean field theory and Kondo lattice}
\author{Young-Kwon Han, Debabrata Ghorai, Taewon Yuk, Sang-Jin Sin}
\affiliation{
  Department of Physics, Hanyang University, \\
  Seoul, 04763, South Korea
}
\emailAdd{youngkwonhan346@gmail.com}
\emailAdd{dghorai123@gmail.com}
\emailAdd{tae1yuk@gmail.com}
\emailAdd{sangjin.sin@gmail.com}
\abstract{
  We first study a non-relativistic field theory model for the Kondo lattice by introducing the Kondo condensation, whose main effect is the hybridization of the flat band of the localized electron with dispersive one of the itinerant electron.
  The problem here is that the resulting Kondo condensation arises only in strong coupling where the validity of the mean field theory is questionable.
  Therefore, we build a holographic mean field theory of the Kondo lattice with strong coupling by identifying the effect of the lattice with the fermion's spectral shape due to the coupling with the order parameter representing the symmetry breaking. 
  For the flat band spectrum we use the mixed quantization,  and for the dispersive spectrum we intoduce the second fermion in standard quantization.
  The coupling of the two fermions with the scalar order representing the Kondo condensation provides the hybrization of the two spectrum, reproducing the main feature of the Kondo lattice together with the fuzzy character of the spectrum of the strongly coupled system. 
 }

\begin{document}

\maketitle
\flushbottom

\section{Introduction}

The Kondo effect describes the scattering of conduction electrons in a metal due to localized magnetic impurities, leading to the minimum of the resistance as a function of temperature.
Since its discovery in 1930 \cite{Meissner1930}, it has been one of the leading motivations in developing modern quantum field theory as well as in the physics of strongly correlated condensed matter.

In 1961, P. W. Anderson introduced the exchange interaction between the localized electron in the $d$-orbital and itinerant electrons in the $s$-orbital to explain the presence of the localized magnetic moment in metals \cite{Anderson1961}.
In 1964, based on the coupling between the spin impurity and itinerant electron introduced in refs. \cite{Zener1951, Ruderman1954, Kasuya1956, Yosida1957_1, Yosida1957_2}, J. Kondo explained the resistance minimum \cite{Kondo1964} using the third-order perturbative field theory of another $s$-$d$ exchange model.
His result is summarized as a logarithmic contribution, $\ln T_c/T$, to the resistivity.
The Kondo temperature, $T_{c}$, is the limit below which his result is not valid, and the divergence problem in the $T \to 0$ limit was resolved by K. Wilson in the 1970s by inventing the modern concept of renormalization group \cite{Wilson:1973jj}.
In the late 1960s, J. R. Schrieffer and his collaborators \cite{Schrieffer1966, Coqblin1969} showed that the Kondo model is a low-energy effective model of the Anderson model using a canonical transformation, which is now called Schrieffer-Wolff transformation, that projects out the high-energy degrees of freedom.
From the modern point of view, because the interaction is marginally relevant, the Kondo model lies in the strong coupling regime of the Anderson impurity model, so the Kondo problem is essentially the physics of a strongly coupled system.
A great deal of research on Kondo systems followed afterward, using the large-$N$ approach, renormalization group, and conformal field theory \cite{Coqblin1969, Read1983_1, Read1983_2, Coleman1983, Coleman1984, Kotliar1986, Coleman1987, Millis1987, Georges1992, Fisk1996, Tsunetsugu1997, Riseborough2000, Dzero2010, Coleman2015, PhysRevB.52.9528}, and the Kondo physics has been the testing ground of new ideas for the non-perturbative or exact calculational method.

When the density of impurities is low, we do not need to consider the impurity-impurity correlation, and we may consider one impurity problem.
For higher density impurities, the problem can be classified into random impurities and impurity lattice.
For the latter, the Kondo effect likely explains the formation of heavy fermions and Kondo insulators.
The main expected characteristic of the Kondo lattice \cite{Auerbach1986, Coleman1983} is the enlarged Fermi surface of heavy fermions formed by the hybridization of itinerant electrons and localized magnetic impurities forming a regular lattice structure.
For the former, the concept of the Kondo condensation, the condensation of the pair of the itinerant electrons and the magnetic impurities in the Kondo lattice, i.e., a magnetic version of the Bardeen-Cooper-Schrieffer pair condensation was suggested, and it was experimentally observed \cite{Im2023}.
Independently, mean field studies on the condensation of the pair of heavy and light fermions have been done to study the Kondo lattice \cite{Yasui2017, Suzuki2017, Yasui2019, Suenaga2020, Araki2021, Ishikawa2021, Suenaga2022, Hattori2023} using a continuum relativistic field theory based on the Nambu-Jona-Lasinio model \cite{Nambu1961_1, Nambu1961_2}.

In this paper, we begin by analyzing a non-relativistic field theory model for the Kondo lattice to simplify the relativistic model \cite{Yasui2017, Suzuki2017, Yasui2019, Suenaga2020, Araki2021, Ishikawa2021, Suenaga2022, Hattori2023} and to consider higher-order corrections effectively.
We show that the Fierz identity and the mean field approximation relate our model to the continuum version of the Anderson model without the Schrieffer-Wolff transformation.
We also find that if only the vector-type $s$-$d$ interaction term exists in our model, then it is unstable.
Hence, we add a scalar-type $s$-$d$ coupling to stabilize the system.
We introduce both scalar and vector types of the Kondo condensation and show that the Kondo condensation forms when the temperature is low and the $s$-$d$ mixing is strong.
The type of the Kondo condensation is determined by the relation between vector- and scalar-type coupling constants.
We focus on the case where only the scalar-type Kondo condensation forms so that our model can be reduced to a lattice version of the Anderson model.
The phase transition in this model is a first-order quantum phase transition.

As we mentioned before, the Kondo condensation forms only if the heavy-light coupling is strong, for which case the mean field theory is not supposed to work.
The subtlety in the usual large-$N$ approaches to justify the mean field approximation is that there is no such number in real materials. 
To resolve these problems, we try to describe the Kondo condensation in the Kondo lattice using the holographic framework \cite{Maldacena1999, Witten1998, Gubser1998, Hartnoll2009, Herzog2009, McGreevy2010, Horowitz2011, Sachdev2011}, which is supposed to handle strong correlations.

In fact, various holographic studies about single-impurity systems have already been performed before.
One approach is in the top-down manner by considering the single impurity described by a D$(8 - p)$-brane, where $p$ is the spatial dimension of the dual gauge theory \cite{Kachru2010, Kachru2011, Muck2011, Faraggi2011, Jensen2011, Karaiskos2012, Harrison2012, Benincasa2012_1, Faraggi2012, Benincasa2012_2, Itsios2013}.
Another one is a bottom-up approach considering the Chern-Simons gauge field in the Ba\~{n}ados-Teitelboim-Zanelli black hole background and the Yang-Mills gauge field in a localized $\text{AdS}_{2}$ subspace, where the information of the single impurity is encoded in the flux of the Yang-Mills field \cite{Erdmenger2013, Bannon2016, Erdmenger2016_1, Erdmenger2016_2, Erdmenger2017_1, Erdmenger2017_2, Erdmenger2017_3, Padhi2018, Erdmenger2020}.
Some similarities between the fractionalized Fermi liquid phase of a lattice-impurity system and the holographic metals were also pointed out \cite{Sachdev2010_1, Sachdev2010_2}.
The ref. \cite{Im2023} suggested a holographic Kondo model of dense random impurities described by the scalar-type Kondo condensation.

Yet, there has not been any holographic Kondo lattice model that gives the fermionic spectrum even within the simplest bottom-up approach.
In fact, the holographic theory is a continuum field theory describing a low-energy effective theory, and introducing lattice eliminates the small spacetime structure, destroying the continuum structure.
This is a challenge.

Recently, we developed the holographic mean field theory \cite{Oh2021_1, Sukrakarn:2023ncp}, where the fermion bilinear is coupled to order parameter fields of various tensor types.
The non-zero order parameter of a particular tensor type corresponds to a specific rotational symmetry breaking.
With various symmetry breaking types, the fermion spectral functions produce features like separated Dirac cones, nodal lines, flat bands as well as gaps of s- p- and d-wave types.
In fact, these exaust important features that can be produced by lattices.
We interpret this as the realization of the lattice in the continuum field theory with symmetry breaking.
In other words, the role of the lattice in the low energy is nothing but a particular type of rotational symmetry breaking.

Extending this idea, we define a holographic Kondo lattice as the holographic mean field theory which produces the spectrum that is expected in the Kondo lattice, that is, a band structure that is a hybridization of a flat band with a hyperbolic one.
The holographic flat band has already been realized in ref. \cite{Laia2011} by the mixed quantization, and it was interpreted \cite{Han2023} as the consequence of cancelation between the two components, which is precisely the mechanism for producing the flat bands by forming the compact localized orbits in lattice models.
The rest of the idea is to introduce another holographic fermion in standard quantization \cite{Cubrovic1993, Lee2009, Iqbal2009, Liu2011, Faulkner2011, Liu2018, Seo2018, Chakrabarti2019, Oh2021_1, Oh2021_2, Yuk2023} and give interaction between the two flavors to hybridize the spectrum.

In summary, after getting intuition from our non-relativistic mean field theory, we construct an effective model for the Kondo lattice in the holographic framework of spinors using the standard-mixed quantization.
Finally, by calculating the fermionic spectral function and the density of states following the standard lore, we show that our proposed holographic Kondo lattice model indeed produces the main features of the Kondo condensation in the Kondo lattice system including the extended Fermi volume as well as the large fermion mass.

In section 2, we study the mean field theory in the non-relativistic framework.
In section 3, we construct a holographic mean field theory for the Kondo lattice and calculate the spectral functions.
We summarize and discuss in section 4.
There are four appendices describing the mathematical details, which are omitted in the main text.

\section{Non-relativistic mean field model for the Kondo condensation}

\subsection{Setup}

Based on refs. \cite{Ruderman1954, Kasuya1956, Yosida1957_1, Yosida1957_2, Kondo1964, Nambu1961_1, Nambu1961_2, Yasui2017, Suzuki2017, Yasui2019, Suenaga2020, Araki2021, Ishikawa2021, Suenaga2022, Hattori2023}, we construct a non-relativistic model for the Kondo lattice in continuum limit as follows:
\begin{equation}
  \begin{split}
    \mathcal{L} = &\ \psi^{\dagger} \left( i \pdv{t} + \frac{\nabla^{2}}{2 m} + \mu \right) \psi + \chi^{\dagger} \left( i \pdv{t} - \lambda \right) \chi \\
                  &\ + \frac{g_{l}}{2} (\psi^{\dagger} \psi)^{2} - g_{s} (\psi^{\dagger} \psi) (\chi^{\dagger} \chi) - g_{v} (\psi^{\dagger} \vec{\sigma} \psi) \cdot (\chi^{\dagger} \vec{\sigma} \chi). \label{equation:starting_lagrangian}
  \end{split}
\end{equation}
$\psi \equiv (\psi_{\uparrow}, \psi_{\downarrow})^{T}$ and $\chi \equiv (\chi_{\uparrow}, \chi_{\downarrow})^{T}$ describe the light and heavy fermions, respectively.
$m$ is the mass of the light fermion.
$\mu$ is the chemical potential for the light fermion.
$\lambda$ is the energy level of the heavy fermion without hybridization.
$\vec{\sigma} = (\sigma_{1}, \sigma_{2}, \sigma_{3})$ are the Pauli matrices.
$g_{l}$ is the light-light coupling constant.
$g_{s}$ and $g_{v}$ are the scalar- and vector-type heavy-light coupling constants, respectively.
Using the Fierz identity, we can write the Lagrangian as
\begin{equation}
  \begin{split}
    \mathcal{L} = &\ \psi^{\dagger} \left( i \pdv{t} + \frac{\nabla^{2}}{2 m} + \mu \right) \psi + \chi^{\dagger} \left( i \pdv{t} - \lambda \right) \chi \\
                  &\ + \frac{g_{l}}{2} (\psi^{\dagger} \psi)^{2} + g_{s}' (\psi^{\dagger} \chi) (\chi^{\dagger} \psi) + g_{v}' (\psi^{\dagger} \vec{\sigma} \chi) \cdot (\chi^{\dagger} \vec{\sigma} \psi),
  \end{split}
\end{equation}
where
\begin{equation}
  g_{s}' \coloneqq \frac{g_{s} + 3 g_{v}}{2}, \quad
  g_{v}' \coloneqq \frac{g_{s} - g_{v}}{2}.
\end{equation}
The Hubbard-Stratonovich transformation with $M$ and the Kondo condensation $\Delta \sim \expval*{\psi^{\dagger} \chi}$,
\begin{align}
  \frac{g_{l}}{2} (\psi^{\dagger} \psi)^{2} \to &\ -\frac{M^{2}}{2 g_{l}} - M (\psi^{\dagger} \psi), \\
  g_{s}' (\psi^{\dagger} \chi) (\chi^{\dagger} \psi) \to &\ -\frac{|\Delta_{s}|^{2}}{g_{s}'} + \Delta_{s} (\chi^{\dagger} \psi) + (\psi^{\dagger} \chi) \Delta_{s}^{*}, \\
  g_{v}' (\psi^{\dagger} \vec{\sigma} \chi) \cdot (\chi^{\dagger} \vec{\sigma} \psi) \to &\ -\frac{|\Delta_{v}|^{2}}{g_{v}'} + \vec{\Delta}_{v} \cdot (\chi^{\dagger} \vec{\sigma} \psi) + (\psi^{\dagger} \vec{\sigma} \chi) \cdot \vec{\Delta}_{v}^{*},
\end{align}
gives the following mean field Lagrangian
\begin{equation}
  \mathcal{L}_{\text{MF}} = \Psi^{\dagger} D \Psi - U,
\end{equation}
where
\begin{align}
  \Psi^{\dagger} \coloneqq &\ \mqty(\psi^{\dagger} & \chi^{\dagger}), \quad \Psi \coloneqq \mqty(\psi \\ \chi), \\
  D \coloneqq &\ \mqty(i \pdv{t} + \frac{\nabla^{2}}{2 m} + \mu - M & \Delta_{s}^{*} + \vec{\sigma} \cdot \vec{\Delta}_{v}^{*} \\ \Delta_{s} + \vec{\sigma} \cdot \vec{\Delta}_{v} & i \pdv{t} - \lambda), \\
  U \coloneqq &\ \frac{M^{2}}{2 g_{l}} + \frac{|\Delta_{s}|^{2}}{g_{s}'} + \frac{|\vec{\Delta}_{v}|^{2}}{g_{v}'}. \label{equation:potential}
\end{align}
The thermodynamic potential is given by (see appendix \ref{appendix:the_thermodynamic_potential_of_the_non-relativistic_mean-field_model} and ref. \cite{Kapusta2006})
\begin{equation}
  \begin{split}
    \Omega = &\ U + \frac{1}{V} \sum_{|\vec{p}| < \Lambda} \sum_{i = 1}^{4} \left\{ -\frac{1}{2} |\omega_{i}(\vec{p})| - \frac{1}{\beta} \ln \left[ 1 + e^{-\beta |\omega_{i}(\vec{p})|} \right] \right\} \\
    = &\ U - \frac{1}{4 \pi^{2}} \int_{0}^{\Lambda} \dd p p^{2} \sum_{i = 1}^{4} |\omega_{i}(p)| - \frac{1}{2 \pi^{2} \beta} \int_{0}^{\Lambda} \dd p p^{2} \sum_{i = 1}^{4} \ln \left[ 1 + e^{-\beta |\omega_{i}(p)|} \right], \label{equation:thermodynamic_potential}
  \end{split}
\end{equation}
where $T \equiv 1 / \beta$ is the temperature, $\Lambda$ is the momentum cutoff, and $\omega_{i}(\vec{p})$ is the energy-momentum dispersion defined by
\begin{align}
  G^{-1}(\omega, \vec{p}) \coloneqq &\ \mqty(\omega - \frac{p^{2}}{2 m} + \mu - M & \Delta_{s}^{*} + \vec{\sigma} \cdot \vec{\Delta}_{v}^{*} \\ \Delta_{s} + \vec{\sigma} \cdot \vec{\Delta}_{v} & \omega - \lambda), \\
  \det G^{-1}(\omega, \vec{p}) \equiv &\ \prod_{i = 1}^{4} [\omega - \omega_{i}(\vec{p})].
\end{align}
In this paper, we set $\Lambda$ to be unity.
The first, second, and third terms in the right-hand side of eq. \eqref{equation:thermodynamic_potential} are potential, vacuum, and thermal contributions to $\Omega$, respectively.

\subsection{Energy-momentum dispersion}

\begin{figure}[t]
  \centering
  \begin{minipage}[c][3.3 in][t]{3.8 in}
    \begin{subfigure}{3.8 in}
      \centering
      \includegraphics[width = 3.1 in]{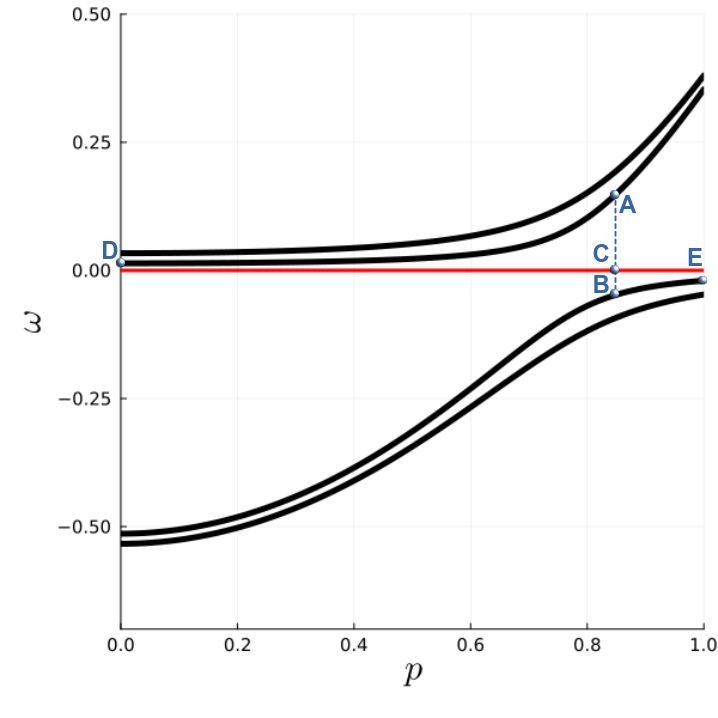}
      \caption{\small $\omega(p)$ with condensation.}
    \end{subfigure}
  \end{minipage}
  \begin{minipage}[c][3.3 in][t]{1.9 in}
    \begin{subfigure}{1.9 in}
      \centering
      \includegraphics[width = 1.4 in]{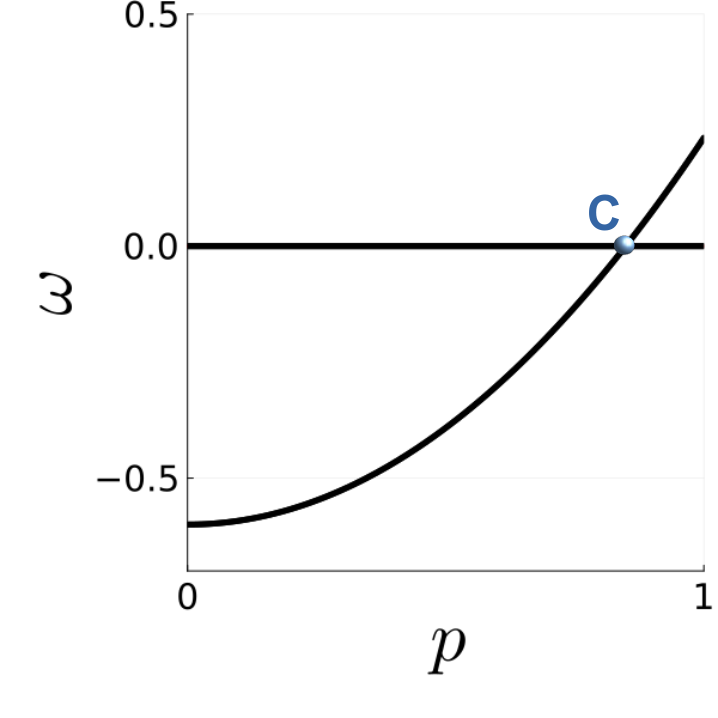}
      \caption{\small $\omega(p)$ without condensation.}
    \end{subfigure}
    \begin{subfigure}{1.9 in}
      \centering
      \includegraphics[width = 1.4 in]{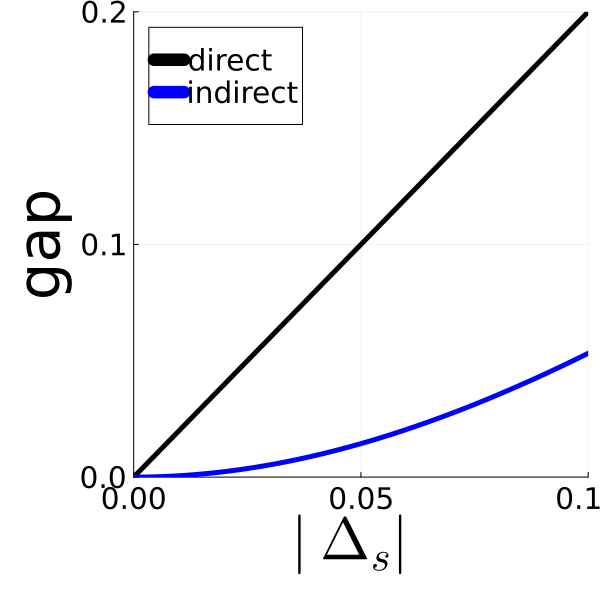}
      \caption{\small Hybridization gap.}
    \end{subfigure}
  \end{minipage}
  \caption{\small
    Energy-momentum dispersion and the hybridization gap.
    $m = \mu = 0.6$, $\lambda = 10^{-8}$.
    (a,b) Black lines represent $\omega_{i}$ and red lines represent the Fermi level.
    We used $M = 0.1$, $|\Delta_{s}| = 0.1$, $|\Delta_{v}| = 0.05$, $\theta = 1$ for (a), while $M = 0$, $|\Delta_{s}| = 0$, $|\Delta_{v}| = 0$, $\theta = 0$ for (b).
    Point C is the band crossing point before the gap opening.
    Point A (B) is the intersection of the vertical line extending from C and the upper (lower) band.
    The gap between A and B is called the direct gap.
    Point D (E) is at the bottom (top) of the upper (lower) band.
    The gap between D and E is called the indirect gap.
    (c) $M = 0$, $|\Delta_{v}| = 0$.
    Black and blue lines show the direct and indirect gaps, respectively.
    The direct gap is approximately linear in $|\Delta_{s}|$, while the indirect gap is quadratic.
  }
  \label{figure:energy-momentum_dispersion_and_the_hybridization_gap}
\end{figure}

Solving $\det G^{-1}(\omega_{i}, \vec{p}) = 0$, we obtain the energy-momentum dispersion
\begin{equation}
  \omega_{i = 1, \cdots, 4} = \mathcal{E}_{+} \pm \sqrt{\mathcal{E}_{-}^{2} + |\Delta_{s}|^{2} + |\vec{\Delta}_{v}|^{2} \pm \sqrt{(|\Delta_{s}|^{2} + |\vec{\Delta}_{v}|^{2})^{2} - |\Delta_{s}^{2} - \vec{\Delta}_{v} \cdot \vec{\Delta}_{v}|^{2}}},
\end{equation}
where
\begin{equation}
  \mathcal{E}_{\pm} \coloneqq \frac{1}{2} \left[ \left( \frac{p^{2}}{2 m} - \mu + M \right) \pm \lambda \right]. \label{equation:dispersion_relation_substitution}
\end{equation}
Since $\omega_{i}$ is invariant under rotation and depends on the phases of the condensations only through their difference, we set an ansatz
\begin{equation}
  \Delta_{s} = |\Delta_{s}|, \quad
  \vec{\Delta}_{v} = |\Delta_{v}| e^{i \theta} \hat{n},
\end{equation}
where $\theta$ is the phase difference between the Kondo condensations and $\hat{n}$ is a unit vector.
Then, we obtain
\begin{equation}
  \omega_{i = 1, \cdots, 4} = \mathcal{E}_{+} \pm \sqrt{\mathcal{E}_{-}^{2} + |\Delta_{s}|^{2} + |\Delta_{v}|^{2} \pm 2 |\Delta_{s}| |\Delta_{v}| \cos \theta}. \label{equation:dispersion_relation}
\end{equation}
$M$ effectively decreases $\mu$, $\omega_{i}$ is invariant under $|\Delta_{s}| \leftrightarrow |\Delta_{v}|$, $|\Delta_{s, v}|$ open the hybridization gap, and the spin degeneracy is broken when $\theta \neq \pm \pi / 2$ and $|\Delta_{s, v}|, \neq 0$ (see eqs. \eqref{equation:dispersion_relation_substitution} and \eqref{equation:dispersion_relation}).
Figure \ref{figure:energy-momentum_dispersion_and_the_hybridization_gap} shows the energy-momentum dispersion and the hybridization gap.

\subsection{Formation of the Kondo condensation}

Given fixed parameters $\underline{\pi} \coloneqq (m, \mu, \lambda, g_{l}, g_{s}, g_{v}, \beta)$, the thermodynamic potential $\Omega$ is a function of four real variables $M$, $|\Delta_{s}|$, $|\Delta_{v}|$, and $\theta$:
\begin{equation}
  \Omega \equiv \Omega_{\underline{\pi}}(M, |\Delta_{s}|, |\Delta_{v}|, \theta).
\end{equation}
Minimizing the function $\Omega_{\underline{\pi}}$, we can obtain $(M, |\Delta_{s}|, |\Delta_{v}|, \theta)$ at thermal equilibrium for given parameters $\underline{\pi}$:
\begin{equation}
  \min \Omega_{\underline{\pi}} \equiv \Omega_{\underline{\pi}}(M_{\text{eq}}(\underline{\pi}), |\Delta_{s}|_{\text{eq}}(\underline{\pi}), |\Delta_{v}|_{\text{eq}}(\underline{\pi}), \theta_{\text{eq}}(\underline{\pi})).
\end{equation}

\begin{figure}[t]
  \centering
  \begin{subfigure}{1.9 in}
    \centering
    \includegraphics[width = 1.9 in]{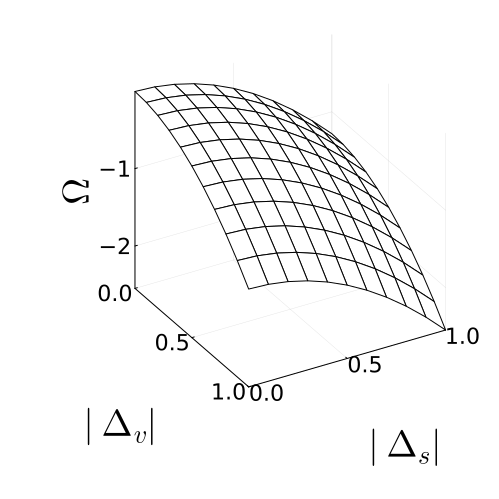}
    \caption{\small Thermodynamic potential with negative coupling.}
    \label{subfigure:thermodynamic_potential_with_negative_coupling}
  \end{subfigure}
  \begin{subfigure}{1.9 in}
    \centering
    \includegraphics[width = 1.7 in]{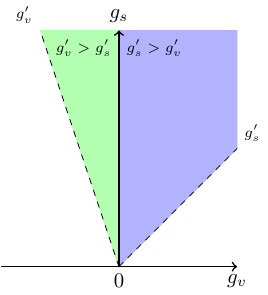}
    \caption{\small Stability condition.}
    \label{subfigure:stability_condition}
  \end{subfigure}
  \begin{subfigure}{1.9 in}
    \centering
    \includegraphics[width = 1.7 in]{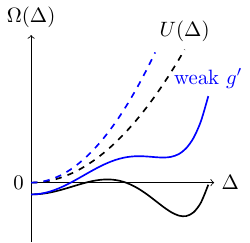}
    \caption{\small Schematic plot of thermodynamic potential.}
    \label{subfigure:schematic_plot_of_thermodynamic_potential}
  \end{subfigure}
  \caption{\small
    (a) $m = \mu = 0.6$, $\lambda = 10^{-8}$, $g_{l} = 0.01$, $g_{s}' = g_{v}' = -0.8$, $T = 0$, $M = 0$, $\theta = \pi / 2$.
    (b) On green (blue) region, $g_{v}' > g_{s}' > 0$ ($g_{s}' > g_{v}' > 0$).
    At the boundary between the green and blue regions, $g_{s}' = g_{v}'$, or equivalently, $g_{v} = 0$.
    On the green and blue regions, the stability condition $g_{s}', g_{v}' > 0$ holds.
    (c) Dashed lines are for $U(\Delta) \sim \Delta^{2} / g'$, solid lines are for $\Omega(\Delta) \sim \Delta^{2} / g' - \frac{1}{2} \int |\omega_{\Delta}|$, and blue lines are for weak $g'$.
  }
\end{figure}

To check the stability of our mean field model, consider $\Omega$ with large condensation $\Delta$:
\begin{equation}
  \begin{split}
    \Omega_{\text{large $\Delta$}} = &\ \underbrace{\left( \frac{M^{2}}{2 g_{l}} + \frac{|\Delta_{s}|^{2}}{g_{s}'} + \frac{|\Delta_{v}|^{2}}{g_{v}'} \right)}_{O(\Delta^{2})} + \underbrace{\frac{1}{V} \sum_{|\vec{p}| < \Lambda} \sum_{i = 1}^{4} \left( -\frac{1}{2} |\omega_{i}(\vec{p})| \right)}_{O(\Delta^{1})} \\
                                     &\ + \underbrace{\frac{1}{V} \sum_{|\vec{p}| < \Lambda} \sum_{i = 1}^{4} \left\{ -\frac{1}{\beta} \ln \left[ 1 + e^{-\beta |\omega_{i}(\vec{p})|} \right] \right\}}_{O(\Delta^{0})}. \label{equation:thermodynamic_potential_with_large_condensation}
  \end{split}
\end{equation}
For large $\Delta$, the potential contribution to $\Omega$, the first term in the right-hand side of eq. \eqref{equation:thermodynamic_potential_with_large_condensation}, is dominant relative to other contributions.
Apart from that, for $\Omega$ to have a minimum with finite $\Delta$, $\Omega$ must not go to negative infinity as $\Delta$ goes to infinity.
Therefore, $g_{l}$, $g_{s}'$, and $g_{v}'$ should be positive so that the potential contribution does not go to negative infinity (see eq. \eqref{equation:potential}).
Figure \ref{subfigure:thermodynamic_potential_with_negative_coupling} shows the thermodynamic potential in the case of $g_{s, v}' < 0$.
We can rewrite the stability condition $g_{s, v}' > 0$ in terms of $g_{s, v}$ in eq. \eqref{equation:starting_lagrangian} as $g_{s} > -3 g_{v}$, $g_{s} > g_{v}$ (see figure \ref{subfigure:stability_condition}).
For the above stability condition to hold, we need nonzero positive $g_{s}$.

\begin{figure}[t]
  \centering
  \begin{subfigure}{1.9 in}
    \centering
    \includegraphics[width = 1.9 in]{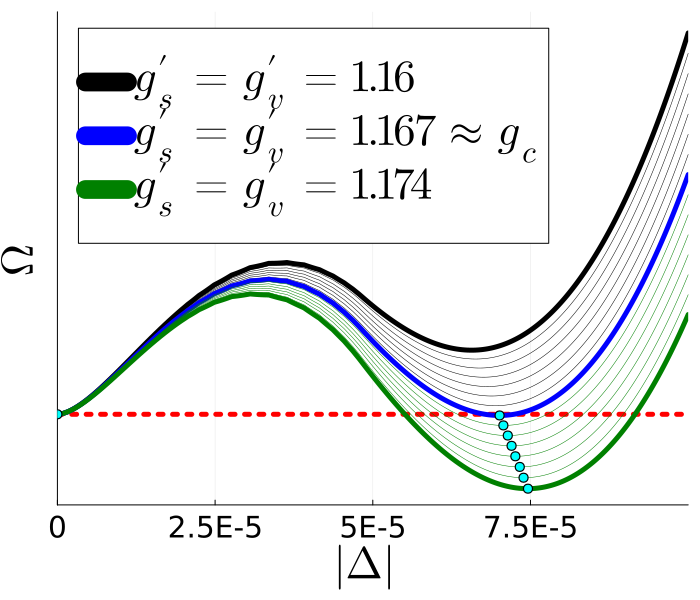}
    \caption{\small $\Omega$ versus $|\Delta|$.}
    \label{subfigure:thermodynamic_potential_at_zero_temperature_with_varying_coupling}
  \end{subfigure}
  \begin{subfigure}{1.9 in}
    \centering
    \includegraphics[width = 1.9 in]{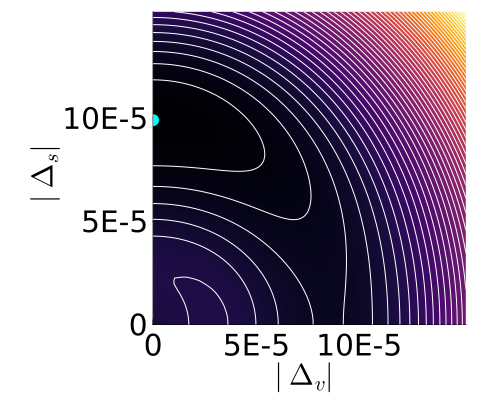}
    \caption{\small $\Omega$ with strong $g_{s}' > g_{v}' > g_{c}$.}
    \label{subfigure:thermodynamic_potential_at_zero_temperature_scalar_case}
  \end{subfigure}
  \begin{subfigure}{1.9 in}
    \centering
    \includegraphics[width = 1.9 in]{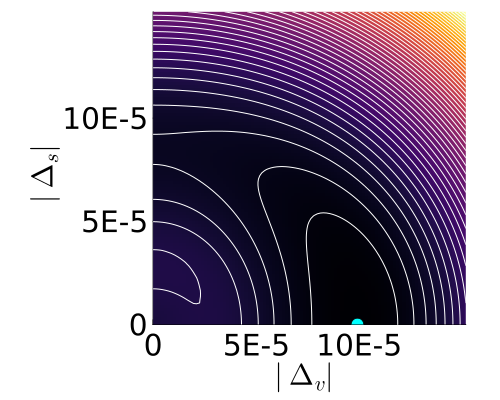}
    \caption{\small $\Omega$ with strong $g_{v}' > g_{s}' > g_{c}$.}
    \label{subfigure:thermodynamic_potential_at_zero_temperature_vector_case}
  \end{subfigure}
  \caption{\small 
    Thermodynamic potential at zero temperature.
    $m = \mu = 0.6$, $\lambda = 10^{-8}$, $g_{l} = 0.01$, $T = 0$, $M = 0 \approx M_{\text{eq}}$, $\theta = \pi / 2 \approx \theta_{\text{eq}}$.
    Cyan points and line represent the minima of $\Omega$.
    (a) shows $\Omega$ versus $|\Delta| \coloneqq \sqrt{|\Delta_{s}|^{2} + |\Delta_{v}|^{2}}$ for various $g_{s}' = g_{v}'$.
    (b,c) The brighter it is, the higher the value of $\Omega$.
    (b) $g_{s}' = 1.21$, $g_{v}' = 1.19$.
    (c) $g_{s}' = 1.19$, $g_{v}' = 1.21$.
  }
  \label{figure:thermodynamic_potential_at_zero_temperature}
\end{figure}

To analyze the effects of the contributions to $\Omega$ and the heavy-light coupling $g'$ at zero temperature qualitatively, consider
\begin{equation}
  \Omega_{T = 0} = \underbrace{\frac{M^{2}}{2 g_{l}} + \frac{|\Delta_{s}|^{2}}{g_{s}'} + \frac{|\Delta_{v}|^{2}}{g_{v}'}}_{\Delta^{2} / g'} + \underbrace{\frac{1}{V} \sum_{|\vec{p}| < \Lambda} \sum_{i = 1}^{4} \left( -\frac{1}{2} |\omega_{i}(\vec{p})| \right)}_{-\frac{1}{2} \int |\omega_{\Delta}|}.
\end{equation}
Without the vacuum contribution $-\frac{1}{2} \int |\omega_{\Delta}|$, minimization of the potential contribution $\Delta^{2} / g'$ gives $\Delta = 0$ (see the dashed lines in figure \ref{subfigure:schematic_plot_of_thermodynamic_potential}).
Therefore, $-\frac{1}{2} \int |\omega_{\Delta}|$ is important to make a new vacuum with nonzero Kondo condensation $\Delta$.
If $g'$ is weak, $\Delta^{2} / g'$ is dominant so that there is no way to make a new vacuum (see the solid blue line in figure \ref{subfigure:schematic_plot_of_thermodynamic_potential}).

Figure \ref{figure:thermodynamic_potential_at_zero_temperature} shows $\Omega$ at $T = 0$ with fixed $M = M_{\text{eq}} = 0$, $\theta = \theta_{\text{eq}} = \pi / 2$.
When $g_{s, v}'$ are weak, the Kondo condensations do not arise (see the black line in figure \ref{subfigure:thermodynamic_potential_at_zero_temperature_with_varying_coupling}).
However, as we increase $g_{s, v}'$, there is a new vacuum, where the Kondo condensation is nonzero (see the blue line in figure \ref{subfigure:thermodynamic_potential_at_zero_temperature_with_varying_coupling}).
Since the thermodynamic potential is invariant under rotation on the $|\Delta_{s}|$-$|\Delta_{v}|$ plane if $g_{s}' = g_{v}'$ and $\theta = \pi / 2$ (see eqs. \eqref{equation:potential}, \eqref{equation:thermodynamic_potential}, and \eqref{equation:dispersion_relation}), there are infinitely many degenerate minima.
The critical coupling constant is $g_{c} \approx 1.167$, and the formation of the Kondo condensation is a first-order quantum phase transition (see figure \ref{subfigure:thermodynamic_potential_at_zero_temperature_with_varying_coupling}).
If we change $g_{s, v}'$ so that $g_{s}' > g_{v}'$, then there is a unique minimum with $|\Delta_{s}| \neq 0$ but $|\Delta_{v}| = 0$ (see figure \ref{subfigure:thermodynamic_potential_at_zero_temperature_scalar_case}).
In the opposite case $g_{s}' < g_{v}'$, only $|\Delta_{v}|$ forms (see figure \ref{subfigure:thermodynamic_potential_at_zero_temperature_vector_case}).
We also perform the finite-temperature calculation in the case of $g_{s}' > g_{v}'$ (see the blue region in figure \ref{subfigure:stability_condition}); and show that, if the temperature is low and the heavy-light coupling is strong, only the scalar-type Kondo condensation forms so that our model is reduced to a Anderson-like model with lattice (see appendix \ref{appendix:numerical_calculation_of_the_non-relativistic_mean-field_model}).

\section{Holographic Kondo lattice model}

\subsection{Setup}

Consider a metric field $g$, a $U(1)$ gauge field $A$, two neutral real scalar fields $\Phi_{\text{s}, \text{ps}}$, and two probe spinor fields $\psi^{(1, 2)}$ in $\text{AdS}_{4}$:
\begin{align}
  S_{\text{tot}} = &\ S_{\text{bg}} + S_{\text{spin}}, \\
  \begin{split}
    S_{\text{bg}} = &\ S_{\text{bg,bdy}} + \int \dd^{4} x \sqrt{-g} \left( R + \frac{6}{L^{2}} - \frac{1}{4} F_{\mu \nu} F^{\mu \nu} \right) \\
                            &\ + \int \dd^{4} x \sqrt{-g} [-(\partial_{\mu} \Phi_{\text{s}}) (\partial^{\mu} \Phi_{\text{s}}) - m_{\text{s}}^{2} \Phi_{\text{s}}^{2} - (\partial_{\mu} \Phi_{\text{ps}}) (\partial^{\mu} \Phi_{\text{ps}}) - m_{\text{ps}}^{2} \Phi_{\text{ps}}^{2}],
  \end{split} \\
  \begin{split}
    S_{\text{spin}} = &\ S_{\text{spin,bdy}} + \sum_{j = 1}^{2} \int \dd^{4} x \sqrt{-g} i \bar{\psi}^{(j)} \left[ \frac{1}{2} \left( \overrightarrow{\slashed{D}}^{(j)} - \overleftarrow{\slashed{D}}^{(j)} \right) - m_{j} \right] \psi^{(j)} \\
                        &\ + \int \dd^{4} x \sqrt{-g} \mqty(\bar{\psi}^{(1)} \\ \bar{\psi}^{(2)})^{T} \mqty(g_{1} \Phi_{\text{ps}} \cdot \Gamma^{5} & V \Phi_{\text{s}} \cdot i \mathbb{I}_{4} \\ V \Phi_{\text{s}} \cdot i \mathbb{I}_{4} & g_{2} \Phi_{\text{s}} \cdot i \mathbb{I}_{4}) \mqty(\psi^{(1)} \\ \psi^{(2)}),
  \end{split} \label{equation:action_spinor_term} \\
  S_{\text{spin,bdy}} = &\ \frac{1}{2} \int \dd^{3} x \sqrt{-h} [\bar{\psi}^{(1)} (i \mathbb{I}_{4}) \psi^{(1)} + \bar{\psi}^{(2)} \Gamma^{\underbar{x} \underbar{y}} \psi^{(2)}], \label{equation:boundary_action_spinor_term}
\end{align}
where
\begin{align}
  \slashed{D}^{(j)} = &\ \Gamma^{a} \tensor{e}{_{a}^{B}} \left( \partial_{B} + \frac{1}{4} \omega_{B c d} \Gamma^{c d} - i q_{j} A_{B} \right), & h = &\ g g^{u u}, \\
  L = &\ 1, & \bar{\psi}^{(j)} = &\ \psi^{(j) \dagger} \Gamma^{\underbar{t}}, \\
  \Gamma^{\underbar{t}} = &\ \sigma_{1} \otimes i \sigma_{2} = \mqty(0 & i \sigma_{2} \\ i \sigma_{2} & 0), & \Gamma^{\underbar{x}} = &\ \sigma_{1} \otimes \sigma_{1} = \mqty(0 & \sigma_{1} \\ \sigma_{1} & 0), \\
  \Gamma^{\underbar{y}} = &\ \sigma_{1} \otimes \sigma_{3} = \mqty(0 & \sigma_{3} \\ \sigma_{3} & 0), & \Gamma^{\underbar{u}} = &\ \sigma_{3} \otimes \sigma_{0} = \mqty(\sigma_{0} & 0 \\ 0 & -\sigma_{0}), \\
  \Gamma^{5} = &\ i \Gamma^{\underbar{t}} \Gamma^{\underbar{x}} \Gamma^{\underbar{y}} \Gamma^{\underbar{u}}, & \Gamma^{a b} = &\ \frac{1}{2} \comm*{\Gamma^{a}}{\Gamma^{b}}.
\end{align}
$m_{\text{s}, \text{ps}}$ and $m_{1, 2}$ are the bulk masses of $\Phi_{\text{s}, \text{ps}}$ and $\psi^{(1, 2)}$, respectively.
We list the motivation for the above action in the following:
\begin{itemize}
\item $g_{1} \bar{\psi}^{(1)} (\Phi_{\text{ps}} \cdot \Gamma^{5}) \psi^{(1)}$ makes a hyperbolic spectrum of the light fermion (to see why we have not chosen the scalar-type interaction, see appendix \ref{appendix:trial_holographic_models_for_the_kondo_lattice}).
\item We consider the standard-mixed quantization to flatten the spectrum of the heavy fermion dual to $\psi^{(2)}$ (see eq. \eqref{equation:boundary_action_spinor_term} and refs. \cite{Li2011, Laia2011, Han2023}).
  The flat spectrum comes from the cancellation of the spinor components making the compact localized states (CLS) \cite{Han2023, Rhim2021}.
\item $g_{2} \bar{\psi}^{(2)} (\Phi_{\text{s}} \cdot i \mathbb{I}_{4}) \psi^{(2)}$ isolates the flat band from others (see appendix \ref{appendix:trial_holographic_models_for_the_kondo_lattice} and ref. \cite{Han2023}).
\item $V \bar{\psi}^{(1)} (\Phi_{\text{s}} \cdot i \mathbb{I}_{4}) \psi^{(2)}$ hybridizes the light and heavy fermions.
\end{itemize}

It turns out that the fermion spectral function depends on the mass of $\Phi_{\text{ps}}$ very sensitively.
Let $\Phi_{\text{ps}} \sim u^{\alpha} + \cdots$ near the boundary.
If $\alpha > 1$ (equivalently if $m_{\text{ps}}^{2} < -2$), then the fermion spectral function has sharp towers of hyperbolic bands, while for $\alpha \le 1$ (equivalently if $m_{\text{ps}}^{2} \ge -2$), then the fermion spectral function has fuzzy filled bands with hyperbolic envelope.
Since we want to imitate the dispersive band like weakly interacting case, we couple the dispersive fermion to the pseudo-scalar with $m_{\text{ps}}^{2} = -9 / 4$.
The spectific value $-9 / 4$ is just to make the power as a simplest possible under the constraint.

\subsection{Background fields}

\begin{figure}[t]
  \centering
  \begin{subfigure}{2.8 in}
    \centering
    \includegraphics[width = 2.8 in]{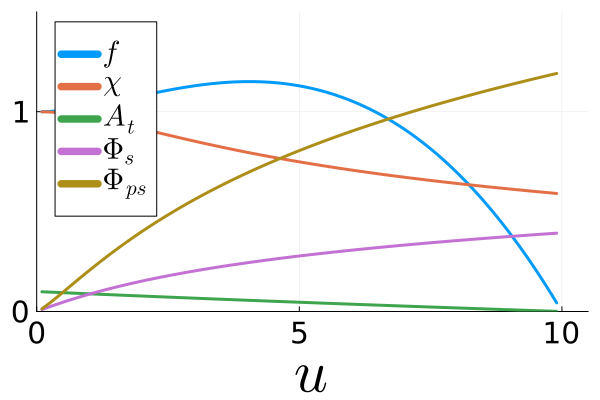}
    \caption{\small At low temperature ($u_{h} = 10, T \approx 0.026$).}
  \end{subfigure}
  \begin{subfigure}{2.8 in}
    \centering
    \includegraphics[width = 2.8 in]{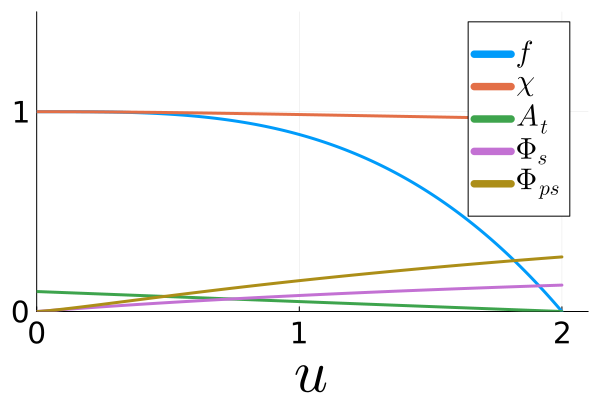}
    \caption{\small At high temperature ($u_{h} = 2, T \approx 0.121$).}
  \end{subfigure}
  \caption{\small Background fields.}
  \label{figure:background_fields}
\end{figure}

We first consider the background fields only by neglecting the probe spinor fields:
\begin{equation}
  \begin{split}
    S_{\text{bg}} = &\ S_{\text{bg,bdy}} + \int \dd^{4} x \sqrt{-g} \left( R + \frac{6}{L^{2}} - \frac{1}{4} F_{\mu \nu} F^{\mu \nu} \right) \\
                    &\ + \int \dd^{4} x \sqrt{-g} [-(\partial_{\mu} \Phi_{\text{s}}) (\partial^{\mu} \Phi_{\text{s}}) - m_{\text{s}}^{2} \Phi_{\text{s}}^{2} - (\partial_{\mu} \Phi_{\text{ps}}) (\partial^{\mu} \Phi_{\text{ps}}) - m_{\text{ps}}^{2} \Phi_{\text{ps}}^{2}].
  \end{split}
\end{equation}
When we choose $m_{\text{s}}^{2} = -2, m_{\text{ps}}^{2} = -9 / 4$ and take ansatz
\begin{align}
  \dd s^{2} = &\ \frac{1}{u^{2}} [-f(u) \chi(u) \dd t^{2} + \dd x^{2} + \dd y^{2}] + \frac{\dd u^{2}}{f(u) u^{2}}, \\
  A = &\ A_{t}(u) \dd t, \quad \Phi_{\text{s}} = \Phi_{\text{s}}(u), \quad \Phi_{\text{ps}} = \Phi_{\text{ps}}(u),
\end{align}
the equations of motion of the background fields $\delta S_{\text{bg}} = 0$ read
\begin{align}
  f' - \left( \frac{3}{u} + \frac{u \Phi_{\text{s}}'^{2}}{2} + \frac{u \Phi_{\text{ps}}'^{2}}{2} \right) f + \frac{3}{u} + \frac{\Phi_{\text{s}}^{2}}{u} + \frac{9 \Phi_{\text{ps}}^{2}}{8 u} - \frac{u^{3} A_{t}'^{2}}{4 \chi} = &\ 0, \\
  \chi' + (u \Phi_{\text{s}}'^{2} + u \Phi_{\text{ps}}'^{2}) \chi = &\ 0, \\
  A_{t}'' - \frac{\chi'}{2 \chi} A_{t}' = &\ 0, \\
  \Phi_{\text{s}}'' + \left( \frac{f'}{f} + \frac{\chi'}{2 \chi} - \frac{2}{u} \right) \Phi_{\text{s}}' + \frac{2}{u^{2} f} \Phi_{\text{s}} = &\ 0, \\
  \Phi_{\text{ps}}'' + \left( \frac{f'}{f} + \frac{\chi'}{2 \chi} - \frac{2}{u} \right) \Phi_{\text{ps}}' + \frac{9}{4 u^{2} f} \Phi_{\text{ps}} = &\ 0,
\end{align}
and the Hawking temperature is given by
\begin{equation}
  T \coloneqq \frac{\sqrt{\chi(u_{h})} |f'(u_{h})|}{4 \pi}.
\end{equation}
The asymptotic behavior near the boundary is as follows:
\begin{align}
  A_{t} \approx &\ \mu - \rho u, \\
  \Phi_{\text{s}} \approx &\ \Phi_{\text{s}, -} u + \Phi_{\text{s}, +} u^{2}, \\
  \Phi_{\text{ps}} \approx &\ \Phi_{\text{ps}, -} u^{3 / 2} \ln u + \Phi_{\text{ps}, +} u^{3 / 2},
\end{align}
where $\mu$ is the chemical potential, $\Phi_{\text{s}/\text{ps}, -}$ are the sources of $\Phi_{\text{s}/\text{ps}}$.
We impose the following boundary condition
\begin{align}
  f(u_{h}) = &\ 0, & \chi(0) = &\ 1, \\
  \mu = &\ 0.1, & A(u_{h}) = &\ 0, \\
  \Phi_{\text{s}, -} = &\ 0.1, & \Phi_{\text{s}}(u_{h}) = &\ \text{finite}, \\
  \Phi_{\text{ps}, -} = &\ 0.1, & \Phi_{\text{ps}}(u_{h}) = &\ \text{finite},
\end{align}
where we have set all control parameters $\mu$, $\Phi_{\text{s}, -}$, and $\Phi_{\text{ps}, -}$ as $0.1$.
In numerical calculation, we cut the domain to be $u \in [\epsilon, u_{h} - \epsilon]$ with some small value $\epsilon$, and then set boundary condition as follows:
\begin{align}
  f(u_{h} - \epsilon) + \epsilon f'(u_{h} - \epsilon) = &\ 0, & \chi(\epsilon) - \epsilon \chi'(\epsilon) - 1 = &\ 0, \\
  A_{t}(\epsilon) - \epsilon A_{t}'(\epsilon) - \mu = &\ 0, & A_{t}(u_{h} - \epsilon) + \epsilon A_{t}'(u_{h} - \epsilon) = &\ 0, \\
  \Phi_{\text{s}}(\epsilon) - \frac{1}{2} \epsilon \Phi_{\text{s}}'(\epsilon) - \frac{1}{2} \Phi_{\text{s}, -} \epsilon = &\ 0, & \epsilon^{2} \Phi_{\text{s}}''(u_{h} - \epsilon) = &\ 0, \\
  \Phi_{\text{ps}}(\epsilon) - \frac{2}{3} \epsilon \Phi_{\text{ps}}'(\epsilon) - \frac{2}{3} \Phi_{\text{ps}, -} \epsilon^{3 / 2} = &\ 0, & \epsilon^{2} \Phi_{\text{ps}}''(u_{h} - \epsilon) = &\ 0.
\end{align}
Figure \ref{figure:background_fields} shows the result of the numerical calculation for the background fields.

\subsection{Probe spinor fields}

\subsubsection{Equations of motion of the probe spinor fields}

After solving the background fields, we consider the probe spinors:
\begin{equation}
  \begin{split}
    S_{\text{spin}} = &\ S_{\text{spin,bdy}} + \sum_{j = 1}^{2} \int \dd^{4} x \sqrt{-g} i \bar{\psi}^{(j)} \left[ \frac{1}{2} \left( \overrightarrow{\slashed{D}}^{(j)} - \overleftarrow{\slashed{D}}^{(j)} \right) - m_{j} \right] \psi^{(j)} \\
                      &\ + \int \dd^{4} x \sqrt{-g} \mqty(\bar{\psi}^{(1)} \\ \bar{\psi}^{(2)})^{T} \mqty(g_{1} \Phi_{\text{ps}} \cdot \Gamma^{5} & V \Phi_{\text{s}} \cdot i \mathbb{I}_{4} \\ V \Phi_{\text{s}} \cdot i \mathbb{I}_{4} & g_{2} \Phi_{\text{s}} \cdot i \mathbb{I}_{4}) \mqty(\psi^{(1)} \\ \psi^{(2)}),
  \end{split}
\end{equation}
with boundary action
\begin{equation}
  S_{\text{spin,bdy}} = \frac{1}{2} \int \dd^{3} x \sqrt{-h} [\bar{\psi}^{(1)} \Gamma^{(1)} \psi^{(1)} + \bar{\psi}^{(2)} \Gamma^{(2)} \psi^{(2)}],
\end{equation}
where $\Gamma^{(j)}$ are some complex $4 \times 4$ matrices that can be written in the block-diagonal form:
\begin{equation}
  \Gamma^{(j)} \equiv \mqty(\gamma_{1 1}^{(j)} & 0 \\ 0 & \gamma_{2 2}^{(j)}).
\end{equation}
Assuming that the background fields are not affected by the probe spinor fields, we have
\begin{equation}
  \begin{split}
    \delta S_{\text{spin}} = &\ (\text{equations of motion term}) \\
    &\ + \sum_{j = 1}^{2} \frac{1}{2} \int \dd^{3} x \sqrt{-h} [\bar{\psi}^{(j)} (\Gamma^{(j)} + i \Gamma^{\underbar{u}}) (\delta \psi^{(j)}) + (\delta \bar{\psi}^{(j)}) (\Gamma^{(j)} - i \Gamma^{\underbar{u}}) \psi^{(j)}],
  \end{split}
\end{equation}
where the equations of motion of the probe spinors are given by
\begin{equation}
  \left[ \mqty(\overrightarrow{\slashed{D}} - m_{1} & 0 \\ 0 & \overrightarrow{\slashed{D}} - m_{2}) - i \mqty(g_{1} \Phi_{\text{ps}} \cdot \Gamma^{5} & V \Phi_{\text{s}} \cdot i \mathbb{I}_{4} \\ V \Phi_{\text{s}} \cdot i \mathbb{I}_{4} & g_{2} \Phi_{\text{s}} \cdot i \mathbb{I}_{4}) \right] \mqty(\psi^{(1)} \\ \psi^{(2)}) = 0.
\end{equation}
Substituting $\psi^{(j)} \eqqcolon (-h)^{-1 / 4} \phi^{(j)}$ to the equations of motion, we get
\begin{align}
  [\Gamma^{a} \tensor{e}{_{a}^{B}} (\partial_{B} - i q_{1} A_{B}) - m_{1} - i g_{1} \Phi_{\text{ps}} \Gamma^{5}] \phi^{(1)} + V \Phi_{\text{s}} \phi^{(2)} = &\ 0, \\
  V \Phi_{\text{s}} \phi^{(1)} + [\Gamma^{a} \tensor{e}{_{a}^{B}} (\partial_{B} - i q_{2} A_{B}) - m_{2} + g_{2} \Phi_{\text{s}}] \phi^{(2)} = &\ 0.
\end{align}
Then, taking
\begin{equation}
  \phi^{(j)}(t, x, y, u) = \int \frac{\dd^{3} k}{(2 \pi)^{3}} e^{-i \omega t + i \vec{k} \cdot \vec{x}} \phi_{k}^{(j)}(u) \quad [k \equiv (\omega, \vec{k}) \equiv (\omega, k_{x}, k_{y})],
\end{equation}
we obtain
\begin{align}
  \partial_{u} \phi_{k}^{(1)} + \Gamma^{\underbar{u}} \left[ \tfrac{-i (\omega + q_{1} A_{t}) \Gamma^{\underbar{t}}}{f \sqrt{\chi}} + \tfrac{i k_{x} \Gamma^{\underbar{x}} + i k_{y} \Gamma^{\underbar{y}}}{\sqrt{f}} + \tfrac{-m_{1} - i g_{1} \Phi_{\text{ps}} \Gamma^{5}}{u \sqrt{f}} \right] \phi_{k}^{(1)} + \tfrac{V \Phi_{\text{s}} \Gamma^{\underbar{u}}}{u \sqrt{f}} \phi_{k}^{(2)} = &\ 0, \\
  \partial_{u} \phi_{k}^{(2)} + \tfrac{V \Phi_{\text{s}} \Gamma^{\underbar{u}}}{u \sqrt{f}} \phi_{k}^{(1)} + \Gamma^{\underbar{u}} \left[ \tfrac{-i (\omega + q_{2} A_{t}) \Gamma^{\underbar{t}}}{f \sqrt{\chi}} + \tfrac{i k_{x} \Gamma^{\underbar{x}} + i k_{y} \Gamma^{\underbar{y}}}{\sqrt{f}} + \tfrac{-m_{2} + g_{2} \Phi_{\text{s}}}{u \sqrt{f}} \right] \phi_{k}^{(2)} = &\ 0.
\end{align}

\subsubsection{Probe spinor fields near the boundary}

The asymptotic behavior of the probe spinor fields near the boundary is as follows:
\begin{equation}
  \phi_{k}^{(j)} \approx \mqty(A_{k}^{(j)} u^{m_{j}} + B_{k}^{(j)} u^{1 - m_{j}} \\ C_{k}^{(j)} u^{1 + m_{j}} + D_{k}^{(j)} u^{-m_{j}}).
\end{equation}
We assume $0 < |m_{j}| < 1 / 2$ so that the leading asymptotic behavior is
\begin{equation}
  \phi_{k}^{(j)} \approx \mqty(A_{k}^{(j)} u^{m_{j}} \\ D_{k}^{(j)} u^{-m_{j}}).
\end{equation}
Define
\begin{align}
  \Gamma_{\pm}^{(j)} \coloneqq &\ \frac{1}{2} (\Gamma^{(j)} \pm i \Gamma^{\underbar{u}}), & \Gamma^{\underbar{t}} \Gamma_{+} \equiv &\ Q^{(j) \dagger} P^{(j)}, \\
  \varphi_{k}^{(j)} \coloneqq &\ \mqty(A_{k}^{(j)} \\ D_{k}^{(j)}), & \bar{\varphi}_{k}^{(j)} \coloneqq &\ \varphi_{k}^{(j)} \Gamma^{\underbar{t}}, \\
  \bar{A}_{k}^{(j)} \coloneqq &\ A_{k}^{(j) \dagger} i \sigma_{2}, & \bar{D}_{k}^{(j)} \coloneqq &\ D_{k}^{(j) \dagger} i \sigma_{2}, \\
  C_{k} \coloneqq &\ \mqty(C_{k}^{(1)} \\ C_{k}^{(2)}) \coloneqq \mqty(Q^{(1)} \varphi_{k}^{(1)} \\ Q^{(2)} \varphi_{k}^{(2)}), & J_{k} \coloneqq &\ \mqty(J_{k}^{(1)} \\ J_{k}^{(2)}) \coloneqq \mqty(P^{(1)} \varphi_{k}^{(1)} \\ P^{(2)} \varphi_{k}^{(2)}), \\
  C_{k} \equiv &\ G_{k} J_{k} \quad (\forall J_{k}),
\end{align}
where $G_{k}$ can be determined by introducing the infalling boundary condition and solving the equations of motion (see appendix \ref{appendix:details_of_the_holographic_kondo_lattice_model}).
Assuming
\begin{equation}
  \rank \Gamma_{\pm}^{(j)} = 2, \quad
  \Gamma_{-}^{(j)} = \Gamma^{\underbar{t}} \Gamma_{+}^{(j) \dagger} \Gamma^{\underbar{t}},
\end{equation}
and substituting the solution to $S_{\text{spin}}$, we have
\begin{equation}
  \begin{split}
    \underbar{S}_{\text{spin}} \coloneqq &\ \eval{S_{\text{spin}}}_{\psi = \psi_{\text{sol}}, u \to 0} \\
    = &\ \sum_{j = 1}^{2} \frac{1}{2} \int \dd^{3} x \frac{\dd^{3} k}{(2 \pi)^{3}} \frac{\dd^{3} k'}{(2 \pi)^{3}} e^{-i (\omega - \omega') t + i (\vec{k} - \vec{k}') \cdot \vec{x}} \eval{\phi_{k}^{(j) \dagger} \Gamma^{\underbar{t}} \Gamma^{(j)} \phi_{k'}^{(j)}}_{\phi = \phi_{\text{sol}}, u \to 0} \\
    = &\ \sum_{j = 1}^{2} \frac{1}{2} \int \frac{\dd^{3} k}{(2 \pi)^{3}} \eval{\mqty(A_{k}^{(j)} u^{m_{j}} \\ D_{k}^{(j)} u^{-m_{j}})^{\dagger} \Gamma^{\underbar{t}} \mqty(\gamma_{1 1}^{(j)} & 0 \\ 0 &  \gamma_{2 2}^{(j)}) \mqty(A_{k}^{(j)} u^{m_{j}} \\ D_{k}^{(j)} u^{-m_{j}})}_{u \to 0} \\
    = &\ \sum_{j = 1}^{2} \frac{1}{2} \int \frac{\dd^{3} k}{(2 \pi)^{3}} (\bar{A}_{k}^{(j)} \gamma_{2 2}^{(j)} D_{k}^{(j)} + \bar{D}_{k}^{(j)} \gamma_{1 1}^{(j)} A_{k}^{(j)}) \\
    = &\ \sum_{j = 1}^{2} \frac{1}{2} \int \frac{\dd^{3} k}{(2 \pi)^{3}} (\bar{\varphi}_{k}^{(j)} \Gamma_{+}^{(j)} \varphi_{k}^{(j)} + \bar{\varphi}_{k}^{(j)} \Gamma^{\underbar{t}} \Gamma_{+}^{(j) \dagger} \Gamma^{\underbar{t}} \varphi_{k}^{(j)}) \\
    = &\ \frac{1}{2} \int \frac{\dd^{3} k}{(2 \pi)^{3}} (C_{k}^{\dagger} J_{k} + J_{k}^{\dagger} C_{k}) \\
    = &\ \frac{1}{2} \int \frac{\dd^{3} k}{(2 \pi)^{3}} (J_{k}^{\dagger} G_{k}^{\dagger} J_{k} + J_{k}^{\dagger} G_{k} J_{k}).
  \end{split}
\end{equation}
Then, using the Gubser-Klebanov-Polyakov-Witten relation \cite{Gubser1998, Witten1998}, we can show that $G_{k}$ is the Green's function of the boundary operator dual to the probe spinor fields \cite{Iqbal2009}.

\subsection{Spectral function and density of states}

After calculating the Green's function $G_{k}^{\text{SM}}$ in the standard-mixed quantization, we consider the spectral function $A(\omega, \vec{k})$ and the density of states $g(\omega)$ that are defined by
\begin{align}
  A(\omega, \vec{k}) \coloneqq &\ \lim_{\delta \to 0^{+}} 2 \Im \tr G_{(\omega + i \delta, \vec{k})}^{\text{SM}}, \\
  g(\omega) \coloneqq &\ \int\limits_{|\vec{k}| < 1} \frac{\dd k_{x} \dd k_{y}}{(2 \pi)^{2}} A(\omega, \vec{k}) = \frac{1}{2 \pi} \int_{0}^{1} \dd k k A(\omega, k),
\end{align}
where we have introduced momentum cutoff $|\vec{k}| < \Lambda = 1$ and used the rotational symmetry of the system.

\begin{figure}[t]
  \centering
  \begin{subfigure}{1.4 in}
    \centering
    \includegraphics[width = 1.4 in]{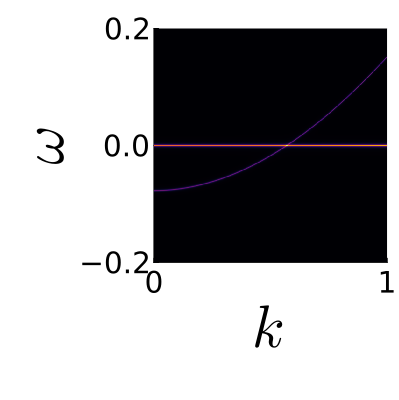}
    \caption{\small $A(\omega, k), V = 0$.}
    \label{subfigure:spectral_function_at_low_temperature_without_hybridization}
  \end{subfigure}
  \begin{subfigure}{1.4 in}
    \centering
    \includegraphics[width = 1.4 in]{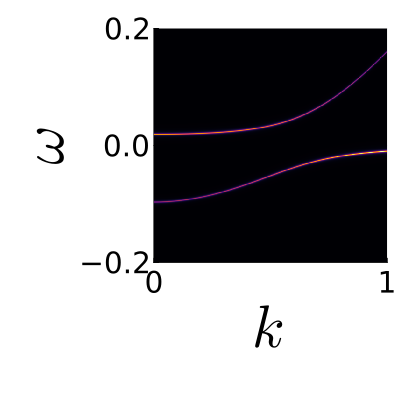}
    \caption{\small $A(\omega, k), V = 0.5$.}
    \label{subfigure:spectral_function_at_low_temperature_with_hybridization}
  \end{subfigure}
  \begin{subfigure}{1.4 in}
    \centering
    \includegraphics[width = 1.4 in]{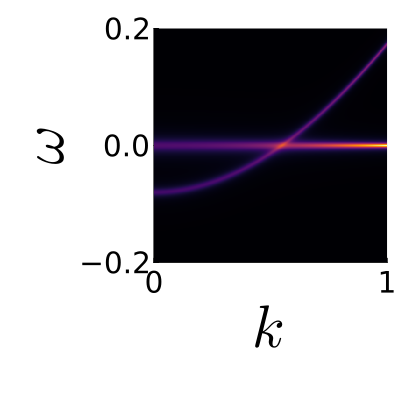}
    \caption{\small $A(\omega, k), V = 0$.}
    \label{subfigure:spectral_function_at_high_temperature_without_hybridization}
  \end{subfigure}
  \begin{subfigure}{1.4 in}
    \centering
    \includegraphics[width = 1.4 in]{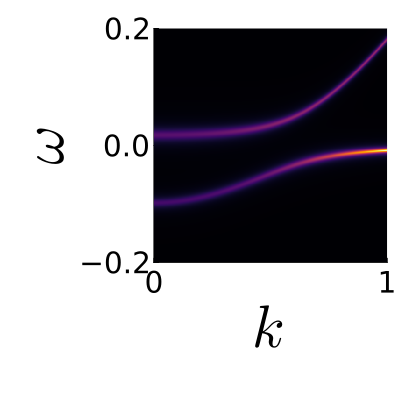}
    \caption{\small $A(\omega, k), V = 0.5$.}
    \label{subfigure:spectral_function_at_high_temperature_with_hybridization}
  \end{subfigure}
  \begin{subfigure}{1.4 in}
    \centering
    \includegraphics[width = 1.4 in]{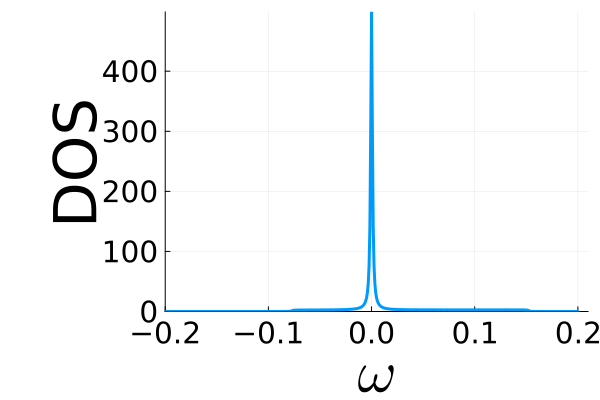}
    \caption{\small $g(\omega), V = 0$.}
    \label{subfigure:density_of_states_at_low_temperature_without_hybridization}
  \end{subfigure}
  \begin{subfigure}{1.4 in}
    \centering
    \includegraphics[width = 1.4 in]{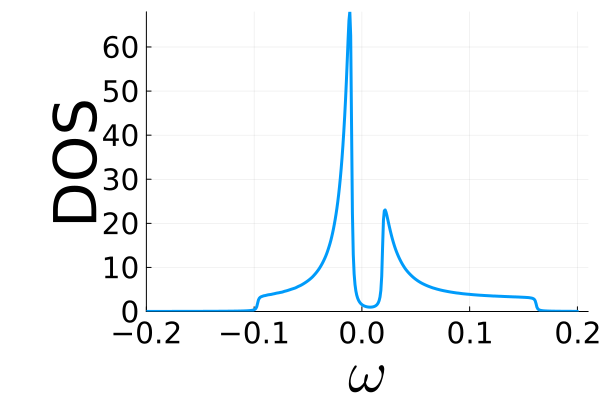}
    \caption{\small $g(\omega), V = 0.5$.}
    \label{subfigure:density_of_states_at_low_temperature_with_hybridization}
  \end{subfigure}
  \begin{subfigure}{1.4 in}
    \centering
    \includegraphics[width = 1.4 in]{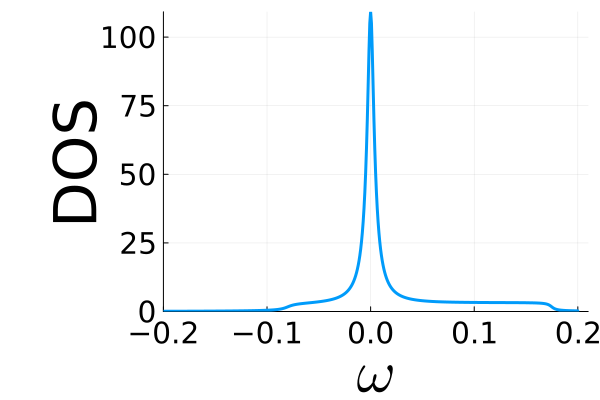}
    \caption{\small $g(\omega), V = 0$.}
    \label{subfigure:density_of_states_at_high_temperature_without_hybridization}
  \end{subfigure}
  \begin{subfigure}{1.4 in}
    \centering
    \includegraphics[width = 1.4 in]{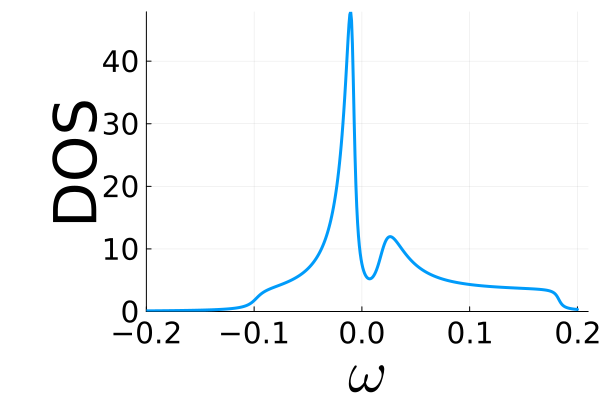}
    \caption{\small $g(\omega), V = 0.5$.}
    \label{subfigure:density_of_states_at_high_temperature_with_hybridization}
  \end{subfigure}
  \caption{\small
    Holographic Kondo lattice model.
    The upper four figures (a,b,c,d) shows the spectral function $A(\omega, k)$ corresponding to the lower four figures (e,f,g,h) of the density of states $g(\omega)$, respectively.
    We used $u_{h} = 10, T \approx 0.026$ for the left four figures (a,b,e,f), while $u_{h} = 2, T \approx 0.121$ for the right four figures (c,d,g,h).
    The brighter points in (a,b,c,d) represent higher values of $A(\omega, k)$.
  }
  \label{figure:holographic_kondo_lattice_model}
\end{figure}

Figure \ref{figure:holographic_kondo_lattice_model} shows the result of numerical calculation with $q_{1} = 23.5$, $q_{2} = 0$, $g_{1} = 10$, $g_{2} = 15$, and $m_{1} = m_{2} = 0^{+}$.
$\delta$ must be $0^{+}$ in principle, but we set $\delta = 10^{-3}$ in the numerical calculation.
Without $V$, there exist hyperbolic and flat spectra independently (see figures \ref{subfigure:spectral_function_at_low_temperature_without_hybridization} and \ref{subfigure:spectral_function_at_high_temperature_without_hybridization}).
As we turn on $V$, the hybridization gap opens (see figures \ref{subfigure:spectral_function_at_low_temperature_with_hybridization} and \ref{subfigure:spectral_function_at_high_temperature_with_hybridization}).
At high temperatures, the spectra spread wide and the hybridization gap becomes a pseudogap, an incomplete gap where the density of states does not touch zero (see figures \ref{subfigure:density_of_states_at_high_temperature_without_hybridization} and \ref{subfigure:density_of_states_at_high_temperature_with_hybridization}).
The spectra and density of states are asymmetric under $\omega \to -\omega$ unlike the holographic Kondo model for random impurities \cite{Im2023}.

\section{Conclusion}

We start from a non-relativistic field theory model with $s$-$d$ interaction based on the Kondo model.
Our model can be transformed into an Anderson-like model using the Fierz identity and the mean field approximation, which is more straightforward than the Schrieffer-Wolff transformation.
We note that we must consider the instability and add a scalar-type $s$-$d$ interaction term to stabilize the mean field calculation.
$M$ shifts the dispersion of the light fermion, while the Kondo condensation hybridizes the light and heavy fermions.
The spin degeneracy is broken when there are two types of Kondo condensation.
Our numerical calculation of the mean field model shows that the Kondo condensation forms at low temperatures if the $s$-$d$ mixing is strong.
The type of the Kondo condensation is determined by the inequality of vector- and scalar-type coupling constants.
If the vector-type Kondo condensation does not arise, our mean field model is reduced to a lattice version of the Anderson model.
The formation of the Kondo condensation at zero temperature with varying $s$-$d$ mixing exemplifies the first-order quantum phase transition.

Then, based on the result of the non-relativistic mean field model, we propose an explicitly-symmetry-broken holographic Kondo lattice model using two-flavor spinors in the standard-mixed quantization.
The spinors in the standard and mixed quantizations give conduction and flat spectra, respectively.
We calculate the fermionic spectral function and density of states and show that the inter-flavor coupling hybridizes the conduction and flat spectra like the Kondo condensation.

There are some of the things we have left out.
We have yet to consider the effect of an external magnetic field, which is crucial to comparing the theoretical result with the experiment in ref. \cite{Im2023}.
On the other hand, the explicitly-symmetry-broken holographic model in this paper cannot explain the formation of the Kondo condensation.
Hence, it would be interesting to study the Kondo lattice model with external fields in strong coupling regions or to construct spontaneous-symmetry-broken holographic Kondo lattice models by including non-linear self-coupling of scalar field or coupling between scalar field with another.
We will consider these issues in future works.

\acknowledgments

This work is supported by Mid-career Researcher Program through the National Research
Foundation of Korea grant No. NRF-2021R1A2B5B02002603, RS-2023-00218998 and NRF-2022H1D3A3A01077468. 
We thank the APCTP for the hospitality during the focus program, where part of this work
was discussed.

\appendix

\section{Thermodynamic potential of the non-relativistic mean field model}
\label{appendix:the_thermodynamic_potential_of_the_non-relativistic_mean-field_model}

The partition function is given by
\begin{equation}
  \mathcal{Z} = \int \mathcal{D} \Psi^{\dagger} \mathcal{D} \Psi \exp \left[ -\int_{0}^{\beta} \dd \tau \int \dd^{3} x (-\Psi^{\dagger} D_{\text{E}} \Psi + U) \right],
\end{equation}
where
\begin{equation}
  D_{\text{E}} := \mqty(-\pdv{\tau} + \frac{\nabla^{2}}{2 m} + \mu - M & \Delta_{s}^{*} + \vec{\sigma} \cdot \vec{\Delta}_{v}^{*} \\ \Delta_{s} + \vec{\sigma} \cdot \vec{\Delta}_{v} & -\pdv{\tau} - \lambda).
\end{equation}
Since $U$ is independent of $\vec{x}$, $\tau$, $\Psi$, and $\Psi^{\dagger}$,
\begin{equation}
  \ln \mathcal{Z} = -\beta V U + \ln \left[ \int \mathcal{D} \Psi^{\dagger} \mathcal{D} \Psi \exp \left( \int_{0}^{\beta} \dd \tau \int \dd^{3} x \Psi^{\dagger} D_{\text{E}} \Psi \right) \right].
\end{equation}
Performing the Fourier transform,
\begin{align}
  \Psi(\tau, \vec{x}) = &\ \sum_{n = -\infty}^{\infty} \sum_{|\vec{p}| < \Lambda} e^{i (\nu_{n} \tau + \vec{p} \cdot \vec{x})} \Psi_{n}(\vec{p}), \\
  \Psi^{\dagger}(\tau, \vec{x}) = &\ \sum_{n = -\infty}^{\infty} \sum_{|\vec{p}| < \Lambda} e^{-i (\nu_{n} \tau + \vec{p} \cdot \vec{x})} \Psi_{n}^{\dagger}(\vec{p}),
\end{align}
where $\nu_{n} \coloneqq (2 n + 1) \pi / \beta$ is the Matsubara frequency, we obtain
\begin{equation}
  \begin{split}
    \ln \mathcal{Z} = &\ -\beta V U + \ln \left\{ \left[ \prod_{n = -\infty}^{\infty} \prod_{|\vec{p}| < \Lambda} V \int \dd \Psi_{n}^{\dagger}(\vec{p}) \dd \Psi_{n}(\vec{p}) \right] \right. \\
            &\ \qquad \qquad \qquad \quad \left. \exp \left[ \beta V \sum_{n = -\infty}^{\infty} \sum_{|\vec{p}| < \Lambda} \Psi_{n}^{\dagger}(\vec{p}) G^{-1}(-i \nu_{n}, \vec{p}) \Psi_{n}(\vec{p}) \right] \right\} \\
    = &\ -\beta V U + \sum_{n = -\infty}^{\infty} \sum_{|\vec{p}| < \Lambda} \ln \left\{ V \int \dd \Psi_{n}^{\dagger}(\vec{p}) \dd \Psi_{n}(\vec{p}) \right. \\
            &\ \qquad \qquad \qquad \qquad \qquad \quad \left. \exp \left[ \beta V \Psi_{n}^{\dagger}(\vec{p}) G^{-1}(-i \nu_{n}, \vec{p}) \Psi_{n}(\vec{p}) \right] \right\} \\ 
    = &\ -\beta V U + \sum_{|\vec{p}| < \Lambda} \sum_{n = -\infty}^{\infty} \ln [\beta^{4} \det G^{-1}(-i \nu_{n}, \vec{p})].
  \end{split}
\end{equation}
Since
\begin{equation}
  \det G^{-1}(\omega, \vec{p}) = \prod_{i = 1}^{4} [\omega - \omega_{i}(\vec{p})],
\end{equation}
we have
\begin{equation}
  \begin{split}
    \ln \mathcal{Z} = &\ -\beta V U + \sum_{|\vec{p}| < \Lambda} \sum_{n = -\infty}^{\infty} \ln \left\{ \beta^{4} \prod_{i = 1}^{4} [-i \nu_{n} - \omega_{i}(\vec{p})] \right\} \\
    = &\ -\beta V U + \frac{1}{2} \sum_{|\vec{p}| < \Lambda} \sum_{i = 1}^{4} \sum_{n = -\infty}^{\infty} \ln [(2 n + 1)^{2} \pi^{2} + \beta^{2} \omega_{i}^{2}(\vec{p})] \\
    = &\ -\beta V U + \frac{1}{2} \sum_{|\vec{p}| < \Lambda} \sum_{i = 1}^{4} \sum_{n = -\infty}^{\infty} \int^{\beta^{2} \omega_{i}^{2}(\vec{p})} \frac{\dd \theta^{2}}{(2 n + 1)^{2} \pi^{2} + \theta^{2}} \\
    = &\ -\beta V U + \frac{1}{2} \sum_{|\vec{p}| < \Lambda} \sum_{i = 1}^{4} \int^{\beta^{2} \omega_{i}^{2}(\vec{p})} \frac{\dd \theta^{2}}{\theta} \left( \frac{1}{2} - \frac{1}{e^{\theta} + 1} \right) \\
    = &\ -\beta V U + \sum_{|\vec{p}| < \Lambda} \sum_{i = 1}^{4} \left\{ \frac{1}{2} \beta |\omega_{i}(\vec{p})| + \ln \left[ 1 + e^{-\beta |\omega_{i}(\vec{p})|} \right] \right\},
  \end{split}
\end{equation}
where we have dropped $\beta$-independent terms and used the following summation formula:
\begin{equation}
  \sum_{n = -\infty}^{\infty} \frac{1}{(n - x) (n - y)} = \frac{\pi (\cot \pi x - \cot \pi y)}{y - x}.
\end{equation}
Then, the thermodynamic potential is given by
\begin{equation}
  \begin{split}
    \Omega = &\ -\frac{\ln \mathcal{Z}}{\beta V} = U + \frac{1}{V} \sum_{|\vec{p}| < \Lambda} \sum_{i = 1}^{4} \left\{ -\frac{1}{2} |\omega_{i}(\vec{p})| - \frac{1}{\beta} \ln \left[ 1 + e^{-\beta |\omega_{i}(\vec{p})|} \right] \right\} \\
    = &\ U - \frac{1}{4 \pi^{2}} \int_{0}^{\Lambda} \dd p p^{2} \sum_{i = 1}^{4} |\omega_{i}(p)| - \frac{1}{2 \pi^{2} \beta} \int_{0}^{\Lambda} \dd p p^{2} \sum_{i = 1}^{4} \ln \left[ 1 + e^{-\beta |\omega_{i}(p)|} \right],
  \end{split}
\end{equation}
which is just a collection of fermionic harmonic oscillators \cite{Nakahara2003}.

The thermodynamic potential at zero temperature is given by
\begin{equation}
  \begin{split}
    \Omega = &\ U + f(0, \sqrt{\max(0, \min(B_{+}, \Lambda^{2}))}, A_{-}, \Delta_{+}) \\
             &\ + f(\sqrt{\min(\Lambda^{2}, \max(B_{+}, 0))}, \Lambda, A_{+}, 0) \\
             &\ + f(0, \sqrt{\max(0, \min(B_{-}, \Lambda^{2}))}, A_{-}, \Delta_{-}) \\
             &\ + f(\sqrt{\min(\Lambda^{2}, \max(B_{-}, 0))}, \Lambda, A_{+}, 0),
  \end{split}
\end{equation}
where
\begin{align}
  \Delta_{\pm} \coloneqq &\ \sqrt{|\Delta_{s}|^{2} + |\Delta_{v}|^{2} \pm 2 |\Delta_{s}| |\Delta_{v}| \cos \theta}, \\
  A_{\pm} \coloneqq &\ 2 m (-\mu + M \pm \lambda), \\
  B_{\pm} \coloneqq &\ 2 m \left( \frac{\Delta_{\pm}^{2}}{\lambda} + \mu - M \right),
\end{align}
\begin{equation}
  \begin{split}
    f(p_{1}, p_{2}, A, \Delta) \coloneqq &\ -\frac{1}{8 \pi^{2} m} \int_{p_{1}}^{p_{2}} \dd p p^{2} \sqrt{(p^{2} + A)^{2} + (4 m \Delta)^{2}} \\
    = &\ \tfrac{p_{1}^{3}}{600 \pi^{2} m} \left[ 15 \sqrt{(A + p_{1}^{2})^{2} + (4 m \Delta)^{2}} \right. \\
                                         &\ \qquad \quad \left. + 6 \sqrt{\tfrac{A^{2} p_{1}^{4}}{A^{2} + (4 m \Delta)^{2}}} F_{1}(\tfrac{5}{2}; \tfrac{1}{2}, \tfrac{1}{2}; \tfrac{7}{2}; -\tfrac{p_{1}^{2}}{A + 4 i m \Delta}, -\tfrac{p_{1}^{2}}{A - 4 i m \Delta}) \right. \\
                                         &\ \qquad \quad \left. + 10 \sqrt{A^{2} + (4 m \Delta)^{2}} F_{1}(\tfrac{3}{2}; \tfrac{1}{2}, \tfrac{1}{2}; \tfrac{5}{2}; -\tfrac{p_{1}^{2}}{A + 4 i m \Delta}, -\tfrac{p_{1}^{2}}{A - 4 i m \Delta}) \right] \\
                                         &\ - \tfrac{p_{2}^{3}}{600 \pi^{2} m} \left[ 15 \sqrt{(A + p_{2}^{2})^{2} + (4 m \Delta)^{2}} \right. \\
                                         &\ \qquad \quad \left. + 6 \sqrt{\tfrac{A^{2} p_{2}^{4}}{A^{2} + (4 m \Delta)^{2}}} F_{1}(\tfrac{5}{2}; \tfrac{1}{2}, \tfrac{1}{2}; \tfrac{7}{2}; -\tfrac{p_{2}^{2}}{A + 4 i m \Delta}, -\tfrac{p_{2}^{2}}{A - 4 i m \Delta}) \right. \\
                                         &\ \qquad \quad \left. + 10 \sqrt{A^{2} + (4 m \Delta)^{2}} F_{1}(\tfrac{3}{2}; \tfrac{1}{2}, \tfrac{1}{2}; \tfrac{5}{2}; -\tfrac{p_{2}^{2}}{A + 4 i m \Delta}, -\tfrac{p_{2}^{2}}{A - 4 i m \Delta}) \right],
  \end{split}
\end{equation}
and $F_{1}$ is the Appell hypergeometric function.

\section{Numerical calculation of the non-relativistic mean field model}
\label{appendix:numerical_calculation_of_the_non-relativistic_mean-field_model}

\begin{figure}[t]
  \centering
  \begin{subfigure}{1.4 in}
    \centering
    \includegraphics[width = 1.4 in]{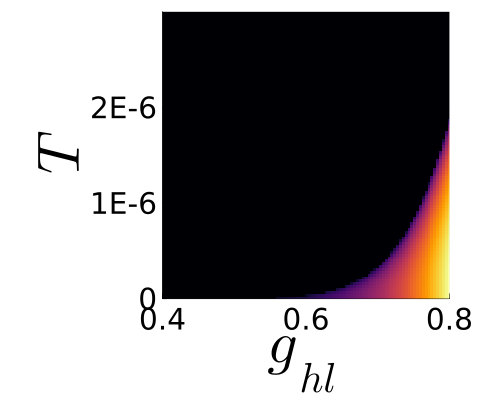}
    \caption{\small $|\Delta_{s}|(g_{hl}, T)$.}
    \label{subfigure:phase_diagram_g_T}
  \end{subfigure}
  \begin{subfigure}{1.4 in}
    \centering
    \includegraphics[width = 1.4 in]{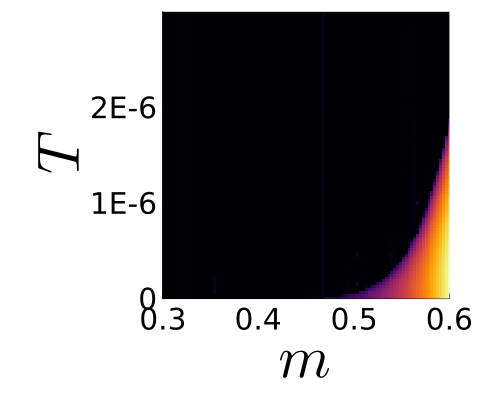}
    \caption{\small $|\Delta_{s}|(m, T)$.}
    \label{subfigure:phase_diagram_m_T}
  \end{subfigure}
  \begin{subfigure}{1.4 in}
    \centering
    \includegraphics[width = 1.4 in]{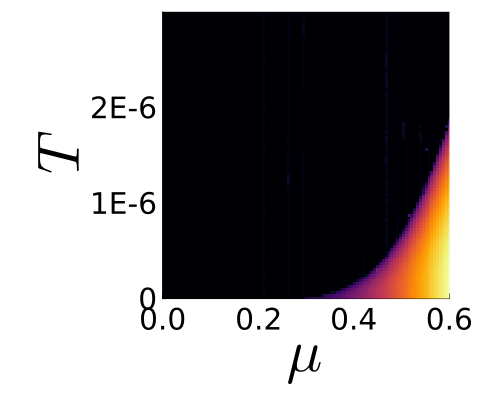}
    \caption{\small $|\Delta_{s}|(\mu, T)$.}
    \label{subfigure:phase_diagram_mu_T}
  \end{subfigure}
  \begin{subfigure}{1.4 in}
    \centering
    \includegraphics[width = 1.4 in]{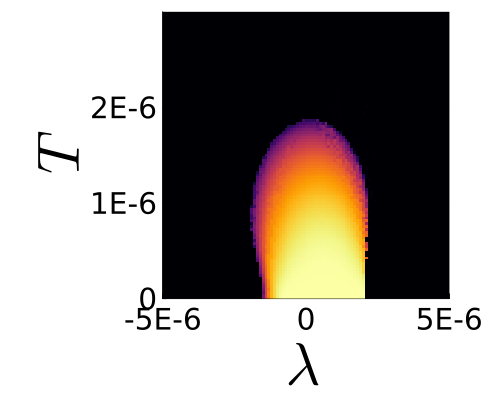}
    \caption{\small $|\Delta_{s}|(\lambda, T)$.}
    \label{subfigure:phase_diagram_lambda_T}
  \end{subfigure}
  \begin{subfigure}{1.4 in}
    \centering
    \includegraphics[width = 1.4 in]{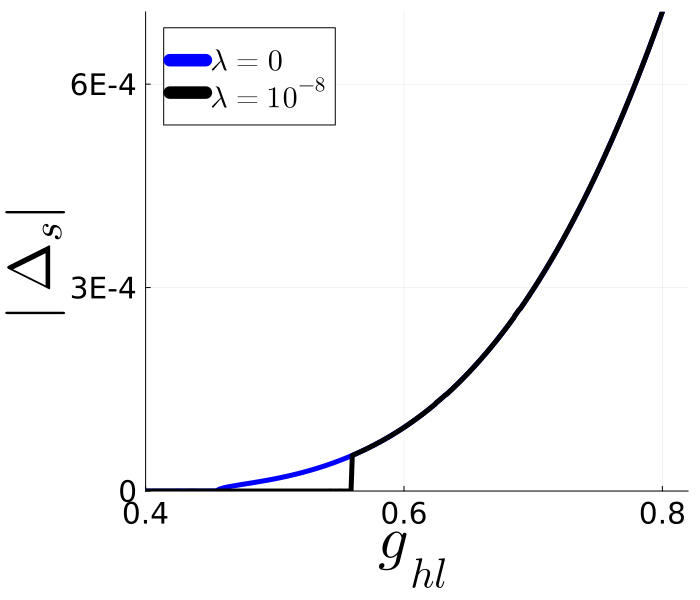}
    \caption{\small $|\Delta_{s}|(g_{h l})$ at $T = 0$.}
    \label{subfigure:gap_equation_g}
  \end{subfigure}
  \begin{subfigure}{1.4 in}
    \centering
    \includegraphics[width = 1.4 in]{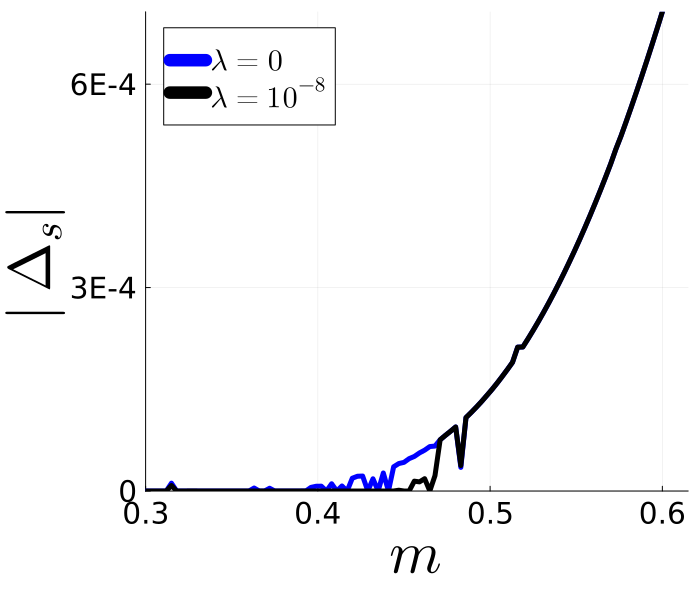}
    \caption{\small $|\Delta_{s}|(m)$ at $T = 0$.}
    \label{subfigure:gap_equation_m}
  \end{subfigure}
  \begin{subfigure}{1.4 in}
    \centering
    \includegraphics[width = 1.4 in]{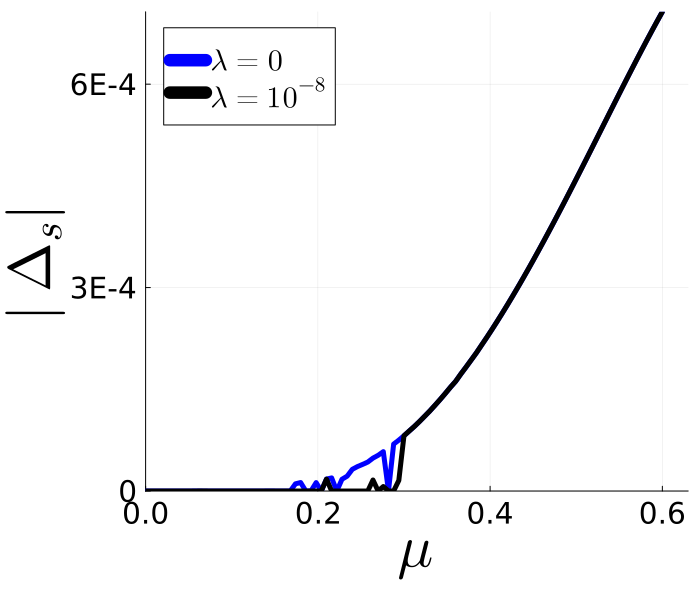}
    \caption{\small $|\Delta_{s}|(\mu)$ at $T = 0$.}
    \label{subfigure:gap_equation_mu}
  \end{subfigure}
  \begin{subfigure}{1.4 in}
    \centering
    \includegraphics[width = 1.4 in]{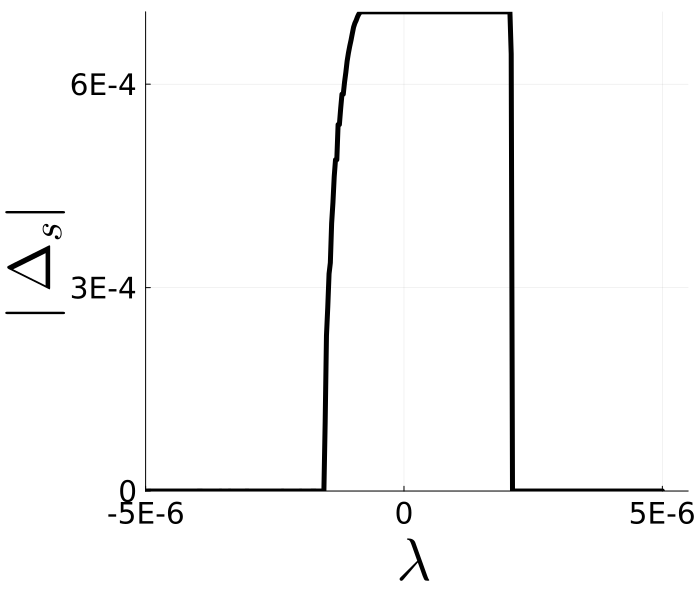}
    \caption{\small $|\Delta_{s}|(\lambda)$ at $T = 0$.}
    \label{subfigure:gap_equation_lambda}
  \end{subfigure}
  \begin{subfigure}{1.4 in}
    \centering
    \includegraphics[width = 1.4 in]{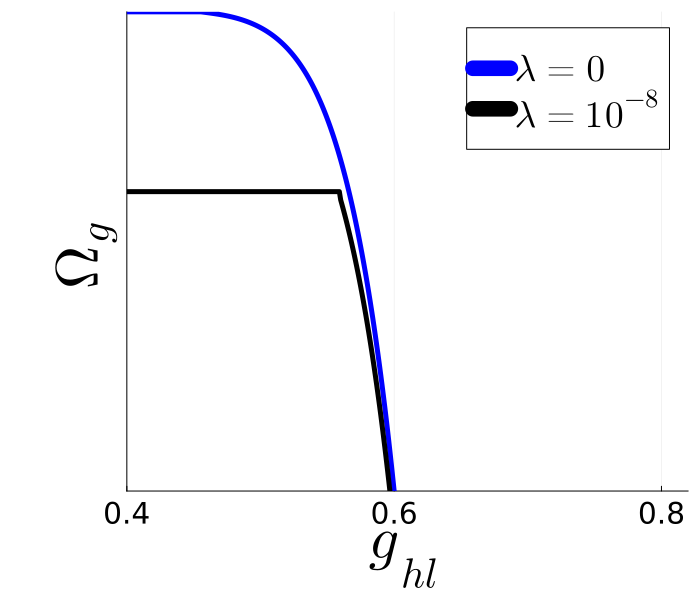}
    \caption{\small $\Omega_{g}(g_{h l})$ at $T = 0$.}
    \label{subfigure:thermodynamic_potential_curve_g}
  \end{subfigure}
  \begin{subfigure}{1.4 in}
    \centering
    \includegraphics[width = 1.4 in]{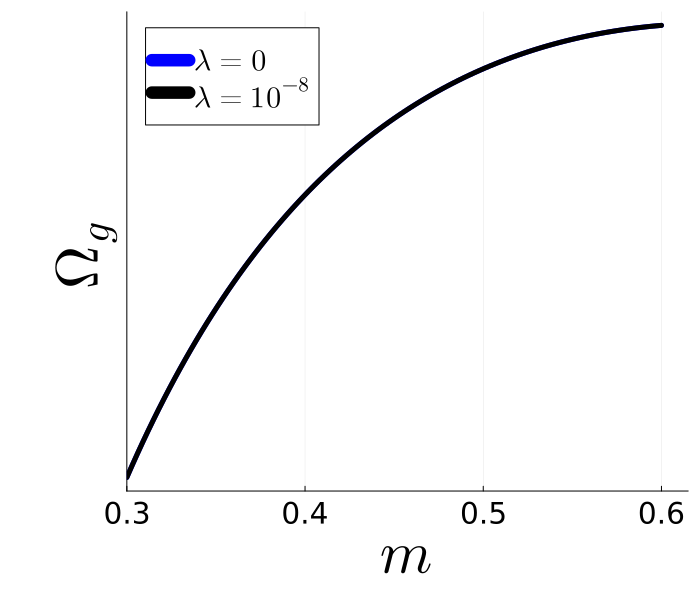}
    \caption{\small $\Omega_{g}(m)$ at $T = 0$.}
    \label{subfigure:thermodynamic_potential_curve_m}
  \end{subfigure}
  \begin{subfigure}{1.4 in}
    \centering
    \includegraphics[width = 1.4 in]{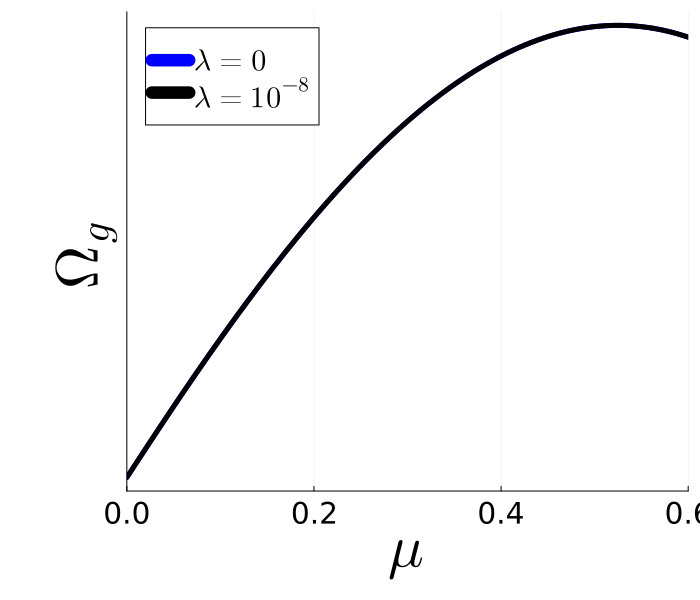}
    \caption{\small $\Omega_{g}(\mu)$ at $T = 0$.}
    \label{subfigure:thermodynamic_potential_curve_mu}
  \end{subfigure}
  \begin{subfigure}{1.4 in}
    \centering
    \includegraphics[width = 1.4 in]{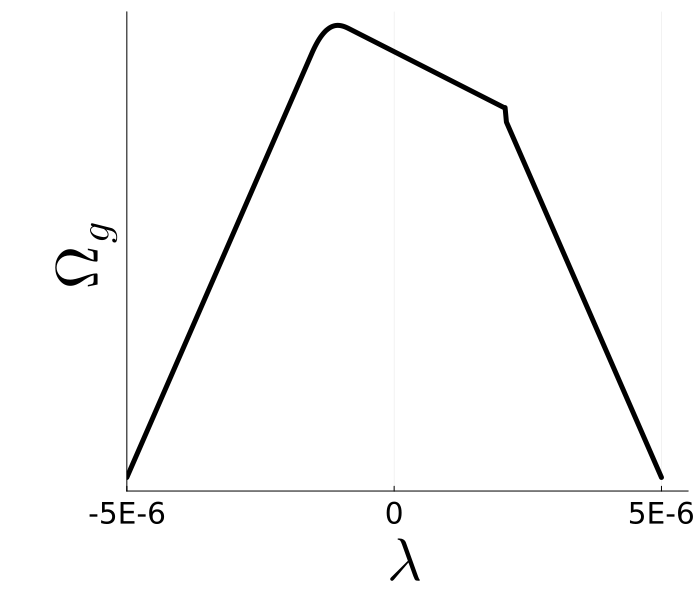}
    \caption{\small $\Omega_{g}(\lambda)$ at $T = 0$.}
    \label{subfigure:thermodynamic_potential_curve_lambda}
  \end{subfigure}
  \caption{\small
    Numerical minimization of the thermodynamic potential.
    $g_{l} = 0.01$, $g_{h l} \coloneqq g_{s}' / 2 = g_{v}'$.
    $\lambda = 0$ for the blue lines.
    (a,e,i) $m = \mu = 0.6$, $\lambda = 10^{-8}$.
    (b,f,j) $g_{hl} = 0.8$, $\mu = 0.6$, $\lambda = 10^{-8}$.
    (c,g,k) $g_{hl} = 0.8$, $m = 0.6$, $\lambda = 10^{-8}$.
    (d,i,l) $g_{hl} = 0.8$, $m = \mu = 0.6$.
    (a,b,c,d) show the phase diagrams.
    The brighter it is, the higher the value of $|\Delta_{s}|$.
    (e,f,g,h) show the the gap equations and (i,j,k,l) show the thermodynamic potential $\Omega_{g}$ of the ground state at $T = 0$ versus $g_{hl}$, $m$, $\mu$, and $\lambda$.
    $M$ and $|\Delta_{v}|$ do not appear in this numerical calculation.
  }
  \label{figure:numerical_minimization_of_the_thermodynamic_potential}
\end{figure}

Figure \ref{figure:numerical_minimization_of_the_thermodynamic_potential} shows the result of the numerical minimization of $\Omega$ with $g_{s}' > g_{v}'$ (see the blue region in figure \ref{subfigure:stability_condition}).
When the temperature is low and the heavy-light coupling is strong, $|\Delta_{s}| \neq 0$, but $M = |\Delta_{v}| = 0$ (see figure \ref{subfigure:phase_diagram_g_T}), so that our model is reduced to a lattice version of the Anderson model as follows:
\begin{equation}
  \mathcal{L}_{\text{MF}} \to \psi^{\dagger} \left( i \pdv{t} + \frac{\nabla^{2}}{2 m} + \mu \right) \psi + \chi^{\dagger} \left( i \pdv{t} - \lambda \right) \chi + |\Delta_{s}| (\psi^{\dagger} \chi + \chi^{\dagger} \psi) - \frac{|\Delta_{s}|^{2}}{g_{s}'}.
\end{equation}
Unlike the mean field theory of superconductivity, where the existence of the condensation is determined only by whether coupling is turned on or off \cite{Altland2010}, in our case, a phase transition occurs at some critical coupling.
When $\lambda = 0$, the phase transition is second-order (see the blue lines in figures \ref{subfigure:gap_equation_g} and \ref{subfigure:thermodynamic_potential_curve_g}).
However, when $\lambda \neq 0$, the phase transition is first-order (see figure \ref{subfigure:thermodynamic_potential_at_zero_temperature_with_varying_coupling} and the black lines in figure \ref{subfigure:gap_equation_g} and \ref{subfigure:thermodynamic_potential_curve_g}).
For first-order quantum phase transition to occur, the system must have a scale \cite{Vojta2003}; in our case, it is $\lambda$.
As we decrease $m$ or $\mu$, $|\Delta_{s}|$ disappears (see figure \ref{subfigure:gap_equation_m} and \ref{subfigure:gap_equation_mu}).
Also, as we increase the magnitude of $\lambda$, $\Delta_{s}$ becomes zero (see figure \ref{subfigure:gap_equation_lambda}).

\section{Details of the holographic Kondo lattice model}
\label{appendix:details_of_the_holographic_kondo_lattice_model}

\subsection{Probe spinor fields in the bulk}

Define
\begin{equation}
  \rho_{k}^{S} \coloneqq \mqty(\phi_{k, 1}^{(1)} \\ \phi_{k, 2}^{(1)} \\ \phi_{k, 1}^{(2)} \\ \phi_{k, 2}^{(2)}), \quad
  \rho_{k}^{C} \coloneqq \mqty(\phi_{k, 3}^{(1)} \\ \phi_{k, 4}^{(1)} \\ \phi_{k, 3}^{(2)} \\ \phi_{k, 4}^{(2)}), \quad
  \varrho_{k}^{S} \coloneqq \mqty(A_{k}^{(1)} \\ A_{k}^{(2)}), \quad
  \varrho_{k}^{C} \coloneqq \mqty(D_{k}^{(1)} \\ D_{k}^{(2)})
\end{equation}
such that
\begin{align}
  \rho_{k}^{S} \approx &\ \mqty(A_{k}^{(1)} u^{m_{1}} \\ A_{k}^{(2)} u^{m_{2}}) = \mqty(u^{m_{1}} & 0 & 0 & 0 \\ 0 & u^{m_{1}} & 0 & 0 \\ 0 & 0 & u^{m_{2}} & 0 \\ 0 & 0 & 0 & u^{m_{2}}) \varrho_{k}^{S}, \\
  \rho_{k}^{C} \approx &\ \mqty(D_{k}^{(1)} u^{-m_{1}} \\ D_{k}^{(2)} u^{-m_{2}}) = \mqty(u^{-m_{1}} & 0 & 0 & 0 \\ 0 & u^{-m_{1}} & 0 & 0 \\ 0 & 0 & u^{-m_{2}} & 0 \\ 0 & 0 & 0 & u^{-m_{2}}) \varrho_{k}^{C}
\end{align}
near the boundary.
Then, the equations of motion of the probe spinor fields read
\begin{equation}
  \partial_{u} \rho_{k}^{S} = \mathcal{D}_{k}^{SS} \rho_{k}^{S} + \mathcal{D}_{k}^{SC} \rho_{k}^{C}, \quad
  \partial_{u} \rho_{k}^{C} = \mathcal{D}_{k}^{CS} \rho_{k}^{S} + \mathcal{D}_{k}^{CC} \rho_{k}^{C},
\end{equation}
where
\begin{align}
  \mathcal{D}_{k}^{SS} \coloneqq &\ \mqty(\frac{m_{1} \sigma_{0}}{u \sqrt{f}} & -\frac{V \Phi_{\text{s}} \sigma_{0}}{u \sqrt{f}} \\ -\frac{V \Phi_{\text{s}} \sigma_{0}}{u \sqrt{f}} & \frac{(m_{2} - g_{2} \Phi_{\text{s}}) \sigma_{0}}{u \sqrt{f}}), \\
  \mathcal{D}_{k}^{SC} \coloneqq &\ \mqty(-\frac{(\omega + q_{1} A_{t}) \sigma_{2}}{f \sqrt{\chi}} - \frac{i k_{x} \sigma_{1} + i k_{y} \sigma_{3}}{\sqrt{f}} + \frac{g_{1} \Phi_{\text{ps}} \sigma_{0}}{u \sqrt{f}} & 0 \\ 0 & -\frac{(\omega + q_{2} A_{t}) \sigma_{2}}{f \sqrt{\chi}} - \frac{i k_{x} \sigma_{1} + i k_{y} \sigma_{3}}{\sqrt{f}}), \\
  \mathcal{D}_{k}^{CS} \coloneqq &\ \mqty(\frac{(\omega + q_{1} A_{t}) \sigma_{2}}{f \sqrt{\chi}} + \frac{i k_{x} \sigma_{1} + i k_{y} \sigma_{3}}{\sqrt{f}} + \frac{g_{1} \Phi_{\text{ps}} \sigma_{0}}{u \sqrt{f}} & 0 \\ 0 & \frac{(\omega + q_{2} A_{t}) \sigma_{2}}{f \sqrt{\chi}} + \frac{i k_{x} \sigma_{1} + i k_{y} \sigma_{3}}{\sqrt{f}}), \\
  \mathcal{D}_{k}^{CC} \coloneqq &\ \mqty(-\frac{m_{1} \sigma_{0}}{u \sqrt{f}} & \frac{V \Phi_{\text{s}} \sigma_{0}}{u \sqrt{f}} \\ \frac{V \Phi_{\text{s}} \sigma_{0}}{u \sqrt{f}} & -\frac{(m_{2} - g_{2} \Phi_{\text{s}}) \sigma_{0}}{u \sqrt{f}}).
\end{align}
For the standard-standard quantization $\Gamma^{(1)} = \Gamma^{(2)} = i \mathbb{I}_{4}$, we can choose
\begin{equation}
  Q^{(1)} = Q^{(2)} = \mqty(0 & \sigma_{1}), \quad P^{(1)} = P^{(2)} = \mqty(-i \sigma_{3} & 0).
\end{equation}
Then,
\begin{align}
  C_{k} = &\ \mqty(Q^{(1)} \varphi_{k}^{(1)} \\ Q^{(2)} \varphi_{k}^{(2)}) = \mqty(\sigma_{1} D_{k}^{(1)} \\ \sigma_{1} D_{k}^{(2)}) = \mqty(\sigma_{1} & 0 \\ 0 & \sigma_{1}) \varrho_{k}^{C}, \\
  J_{k} = &\ \mqty(P^{(1)} \varphi_{k}^{(1)} \\ P^{(2)} \varphi_{k}^{(2)}) = \mqty(-i \sigma_{3} A_{k}^{(1)} \\ -i \sigma_{3} A_{k}^{(2)}) = \mqty(-i \sigma_{3} & 0 \\ 0 & -i \sigma_{3}) \varrho_{k}^{S}; \\
  \therefore \varrho_{k}^{C} = &\ \mqty(\sigma_{1} & 0 \\ 0 & \sigma_{1})^{-1} G_{k} \mqty(-i \sigma_{3} & 0 \\ 0 & -i \sigma_{3}) \varrho_{k}^{S} \quad (\forall \varrho_{k}^{S} \in \mathbb{C}^{4}).
\end{align}
Correspondingly, define
\begin{equation}
  \rho_{k}^{C} \equiv \mqty(\sigma_{1} & 0 \\ 0 & \sigma_{1})^{-1} \mathcal{G}_{k} \mqty(-i \sigma_{3} & 0 \\ 0 & -i \sigma_{3}) \rho_{k}^{S}, \quad
  \tilde{\mathcal{G}}_{k} \coloneqq \mqty(\sigma_{1} & 0 \\ 0 & \sigma_{1})^{-1} \mathcal{G}_{k} \mqty(-i \sigma_{3} & 0 \\ 0 & -i \sigma_{3})
\end{equation}
such that
\begin{equation}
  G_{k} = \lim_{u \to 0} \mqty(u^{2 m_{1}} & 0 & 0 & 0 \\ 0 & u^{2 m_{1}} & 0 & 0 \\ 0 & 0 & u^{2 m_{2}} & 0 \\ 0 & 0 & 0 & u^{2 m_{2}}) \mathcal{G}_{k}.
\end{equation}
From the equations of motion of the probe spinor fields and arbitrariness of $\varrho_{k}^{S}$, we can obtain the flow equations as follows (see ref. \cite{Yuk2023}):
\begin{align}
  &\ & \partial_{u} \rho_{k}^{C} = &\ (\partial_{u} \tilde{\mathcal{G}}_{k}) \rho_{k}^{S} + \tilde{\mathcal{G}}_{k} (\partial_{u} \rho_{k}^{S}) \\
  \Rightarrow &\ & \mathcal{D}_{k}^{CS} \rho_{k}^{C} + \mathcal{D}_{k}^{CC} \rho_{k}^{C} = &\ (\partial_{u} \tilde{\mathcal{G}}_{k}) \rho_{k}^{S} + \tilde{\mathcal{G}}_{k} (\mathcal{D}_{k}^{SS} \rho_{k}^{S} + \mathcal{D}_{k}^{SC} \rho_{k}^{C}) \\
  \Rightarrow &\ & (\mathcal{D}_{k}^{CC} - \tilde{\mathcal{G}}_{k} \mathcal{D}_{k}^{S C}) \tilde{\mathcal{G}}_{k} \rho_{k}^{S} = &\ (\partial_{u} \tilde{\mathcal{G}}_{k} + \tilde{\mathcal{G}}_{k} \mathcal{D}_{k}^{SS} - \mathcal{D}_{k}^{CS}) \rho_{k}^{S}; \\
  \therefore &\ & \partial_{u} \tilde{\mathcal{G}}_{k} = &\ \mathcal{D}_{k}^{CS}  + \mathcal{D}_{k}^{CC} \tilde{\mathcal{G}}_{k} - \tilde{\mathcal{G}}_{k} \mathcal{D}_{k}^{SS} - \tilde{\mathcal{G}}_{k} \mathcal{D}_{k}^{SC} \tilde{\mathcal{G}}_{k},
\end{align}
or equivalently,
\begin{equation}
  \partial_{u} \mathcal{G}_{k} = \tilde{\mathcal{D}}_{k}^{1} + \tilde{\mathcal{D}}_{k}^{2} \mathcal{G}_{k} - \mathcal{G}_{k} \tilde{\mathcal{D}}_{k}^{3} - \mathcal{G}_{k} \tilde{\mathcal{D}}_{k}^{4} \mathcal{G}_{k},
\end{equation}
where
\begin{align}
  \begin{split}
    \tilde{\mathcal{D}}_{k}^{1} \coloneqq &\ \mqty(\sigma_{1} & 0 \\ 0 & \sigma_{1}) \mathcal{D}_{k}^{CS} \mqty(-i \sigma_{3} & 0 \\ 0 & -i \sigma_{3})^{-1} \\
    = &\ \mqty(-\frac{(\omega + q_{1} A_{t}) \sigma_{0}}{f \sqrt{\chi}} - \frac{k_{x} \sigma_{3} + k_{y} \sigma_{1}}{\sqrt{f}} + \frac{g_{1} \Phi_{\text{ps}} \sigma_{2}}{u \sqrt{f}} & 0 \\ 0 & -\frac{(\omega + q_{2} A_{t}) \sigma_{0}}{f \sqrt{\chi}} - \frac{k_{x} \sigma_{3} + k_{y} \sigma_{1}}{\sqrt{f}}),
  \end{split} \label{equation:differential_equation_matrix_1} \\
  \tilde{\mathcal{D}}_{k}^{2} \coloneqq &\ \mqty(\sigma_{1} & 0 \\ 0 & \sigma_{1}) \mathcal{D}_{k}^{CC} \mqty(\sigma_{1} & 0 \\ 0 & \sigma_{1})^{-1} = \mqty(-\frac{m_{1} \sigma_{0}}{u \sqrt{f}} & \frac{V \Phi_{\text{s}} \sigma_{0}}{u \sqrt{f}} \\ \frac{V \Phi_{\text{s}} \sigma_{0}}{u \sqrt{f}} & -\frac{(m_{2} - g_{2} \Phi_{\text{s}}) \sigma_{0}}{u \sqrt{f}}), \label{equation:differential_equation_matrix_2} \\
  \tilde{\mathcal{D}}_{k}^{3} \coloneqq &\ \mqty(-i \sigma_{3} & 0 \\ 0 & -i \sigma_{3}) \mathcal{D}_{k}^{SS} \mqty(-i \sigma_{3} & 0 \\ 0 & -i \sigma_{3})^{-1} = \mqty(\frac{m_{1} \sigma_{0}}{u \sqrt{f}} & -\frac{V \Phi_{\text{s}} \sigma_{0}}{u \sqrt{f}} \\ -\frac{V \Phi_{\text{s}} \sigma_{0}}{u \sqrt{f}} & \frac{(m_{2} - g_{2} \Phi_{\text{s}}) \sigma_{0}}{u \sqrt{f}}), \label{equation:differential_equation_matrix_3} \\
  \begin{split}
    \tilde{\mathcal{D}}_{k}^{4} \coloneqq &\ \mqty(-i \sigma_{3} & 0 \\ 0 & -i \sigma_{3}) \mathcal{D}_{k}^{SC} \mqty(\sigma_{1} & 0 \\ 0 & \sigma_{1})^{-1} \\
    = &\ \mqty(\frac{(\omega + q_{1} A_{t}) \sigma_{0}}{f \sqrt{\chi}} - \frac{k_{x} \sigma_{3} + k_{y} \sigma_{1}}{\sqrt{f}} + \frac{g_{1} \Phi_{\text{ps}} \sigma_{2}}{u \sqrt{f}} & 0 \\ 0 & \frac{(\omega + q_{2} A_{t}) \sigma_{0}}{f \sqrt{\chi}} - \frac{k_{x} \sigma_{3} + k_{y} \sigma_{1}}{\sqrt{f}}).
  \end{split} \label{equation:differential_equation_matrix_4}
\end{align}

\subsection{Probe spinor fields near the horizon}

The asymptotic behavior of the probe spinor fields near the horizon is as follows:
\begin{equation}
  \phi_{k}^{(j)} \approx \underbrace{\left( 1 - \frac{u}{u_{h}} \right)^{-\frac{i \omega}{|f'_{h}| \sqrt{\chi_{h}}}} \mqty(a_{k}^{(j)} \\ b_{k}^{(j)} \\ -b_{k}^{(j)} \\ a_{k}^{(j)})}_{\text{infalling}} + \underbrace{\left( 1 - \frac{u}{u_{h}} \right)^{\frac{i \omega}{|f'_{h}| \sqrt{\chi_{h}}}} \mqty(c_{k}^{(j)} \\ d_{k}^{(j)} \\ d_{k}^{(j)} \\ -c_{k}^{(j)})}_{\text{outgoing}},
\end{equation}
where $f'_{h} \coloneqq f'(u_{h})$, $\chi_{h} \coloneqq \chi(u_{h})$, and $a_{k}^{(j)}, b_{k}^{(j)}, c_{k}^{(j)}, d_{k}^{(j)}$ are expansion coefficients.
Imposing the infalling boundary condition $c_{k}^{(j)} = d_{k}^{(j)} = 0$, we have
\begin{equation}
  \rho_{k}^{S} \approx \left( 1 - \frac{u}{u_{h}} \right)^{-\frac{i \omega}{|f_{h}'| \sqrt{\chi_{h}}}} \mqty(a_{k}^{(1)} \\ b_{k}^{(1)} \\ a_{k}^{(2)} \\ b_{k}^{(2)}), \quad
  \rho_{k}^{C} \approx \left( 1 - \frac{u}{u_{h}} \right)^{-\frac{i \omega}{|f_{h}'| \sqrt{\chi_{h}}}} \mqty(-b_{k}^{(1)} \\ a_{k}^{(1)} \\ -b_{k}^{(2)} \\ a_{k}^{(2)}).
\end{equation}
Therefore, the infalling boundary condition gives
\begin{equation}
  \mqty(-b_{k}^{(1)} \\ a_{k}^{(1)} \\ -b_{k}^{(2)} \\ a_{k}^{(2)}) = \mqty(\sigma_{1} & 0 \\ 0 & \sigma_{1})^{-1} \mathcal{G}_{k}(u_{h}) \mqty(-i \sigma_{3} & 0 \\ 0 & -i \sigma_{3}) \mqty(a_{k}^{(1)} \\ b_{k}^{(1)} \\ a_{k}^{(2)} \\ b_{k}^{(2)}) \Rightarrow \mathcal{G}_{k}(u_{h}) = i \mathbb{I}_{4},
\end{equation}
where we have used arbitrariness of $a_{k}^{(j)}, b_{k}^{(j)}$.

\subsection{From standard-standard to standard-mixed quantization}

Suppose we know the Green's function $G_{k}^{\text{SS}}$ in the standard-standard quantization $\Gamma^{(1)} = \Gamma^{(2)} = i \mathbb{I}_{4}$ (to see how to get $G_{k}^{\text{SS}}$, see appendix \ref{appendix:details_of_the_holographic_kondo_lattice_model}).
When we consider the standard-mixed quantization $\Gamma^{(1)} = i \mathbb{I}_{4}, \Gamma^{(2)} = \Gamma^{\underbar{x} \underbar{y}}$, we can choose
\begin{align}
  P^{(1)} = &\ \mqty(-i & 0 & 0 & 0 \\ 0 & i & 0 & 0), & Q^{(1)} = &\ \mqty(0 & 0 & 0 & 1 \\ 0 & 0 & 1 & 0), \\
  P^{(2)} = &\ \frac{1}{\sqrt{2}} \mqty(-i & 1 & 0 & 0 \\ 0 & 0 & i & 1), & Q^{(2)} = &\ \frac{1}{\sqrt{2}} \mqty(0 & 0 & -i & 1 \\ i & 1 & 0 & 0).
\end{align}
Substituting
\begin{equation}
  \begin{split}
    &\ \varrho_{k}^{C} = \mqty(\sigma_{1} & 0 \\ 0 & \sigma_{1})^{-1} G_{k}^{\text{SS}} \mqty(-i \sigma_{3} & 0 \\ 0 & -i \sigma_{3}) \varrho_{k}^{S} \\
    \Leftrightarrow &\ \mqty(D_{k}^{(1)} \\ D_{k}^{(2)}) = -i \mqty(\sigma_{1} & 0 \\ 0 & \sigma_{1}) G_{k}^{\text{SS}} \mqty(\sigma_{3} & 0 \\ 0 & \sigma_{3}) \mqty(A_{k}^{(1)} \\ A_{k}^{(2)})
  \end{split}
\end{equation}
to
\begin{equation}
  \begin{split}
    C_{k} = G_{k}^{\text{SM}} J_{k} \Leftrightarrow &\ \mqty(Q^{(1)} \varphi_{k}^{(1)} \\ Q^{(2)} \varphi_{k}^{(2)}) = G_{k}^{\text{SM}} \mqty(P^{(1)} \varphi_{k}^{(1)} \\ P^{(2)} \varphi_{k}^{(2)}) \\
    \Leftrightarrow &\ \mqty(Q^{(1)} \mqty(A_{k}^{(1)} \\ D_{k}^{(1)}) \\ Q^{(2)} \mqty(A_{k}^{(2)} \\ D_{k}^{(2)})) = G_{k}^{\text{SM}} \mqty(P^{(1)} \mqty(A_{k}^{(1)} \\ D_{k}^{(1)}) \\ P^{(2)} \mqty(A_{k}^{(2)} \\ D_{k}^{(2)})),
  \end{split}
\end{equation}
we obtain
\begin{equation}
  \mqty(Q^{(1)} \mqty(A_{k}^{(1)} \\ -i \smqty(\sigma_{1} & 0) G_{k}^{\text{SS}} \smqty(\sigma_{3} & 0 \\ 0 & \sigma_{3}) A_{k}^{(1)}) \\ Q^{(2)} \mqty(A_{k}^{(2)} \\ -i \smqty(0 & \sigma_{1}) G_{k}^{\text{SS}} \smqty(\sigma_{3} & 0 \\ 0 & \sigma_{3}) A_{k}^{(2)})) = G_{k}^{\text{SM}} \mqty(P^{(1)} \mqty(A_{k}^{(1)} \\ -i \smqty(\sigma_{1} & 0) G_{k}^{\text{SS}} \smqty(\sigma_{3} & 0 \\ 0 & \sigma_{3}) A_{k}^{(1)}) \\ P^{(2)} \mqty(A_{k}^{(2)} \\ -i \smqty(0 & \sigma_{1}) G_{k}^{\text{SS}} \smqty(\sigma_{3} & 0 \\ 0 & \sigma_{3}) A_{k}^{(2)}));
\end{equation}
\begin{equation}
  \therefore G_{k}^{\text{SM}} = \mqty(Q^{(1)} \mqty(\smqty(\sigma_{0} & 0) \\ -i \smqty(\sigma_{1} & 0) G_{k}^{\text{SS}} \smqty(\sigma_{3} & 0 \\ 0 & \sigma_{3})) \\ Q^{(2)} \mqty(\smqty(0 & \sigma_{0}) \\ -i \smqty(0 & \sigma_{1}) G_{k}^{\text{SS}} \smqty(\sigma_{3} & 0 \\ 0 & \sigma_{3}))) \mqty(P^{(1)} \mqty(\smqty(\sigma_{0} & 0) \\ -i \smqty(\sigma_{1} & 0) G_{k}^{\text{SS}} \smqty(\sigma_{3} & 0 \\ 0 & \sigma_{3})) \\ P^{(2)} \mqty(\smqty(0 & \sigma_{0}) \\ -i \smqty(0 & \sigma_{1}) G_{k}^{\text{SS}} \smqty(\sigma_{3} & 0 \\ 0 & \sigma_{3})))^{-1},
\end{equation}
where we have used arbitrariness of $A_{k}^{(j)}$.
The trace of the Green's function $G_{k}^{\text{SM}}$ in the standard-mixed quantization is then given by
\begin{equation}
  \begin{split}
    \tr G_{k}^{\text{SM}} = &\ (G^{\text{SS}}_{k, 1 1} + G^{\text{SS}}_{k, 2 2}) + \frac{2 (G^{\text{SS}}_{k, 3 3} G^{\text{SS}}_{k, 4 4} - G^{\text{SS}}_{k, 3 4} G^{\text{SS}}_{k, 4 3} - 1)}{G^{\text{SS}}_{k, 3 3} - i G^{\text{SS}}_{k, 3 4} + i G^{\text{SS}}_{k, 4 3} + G^{\text{SS}}_{k, 4 4}} \\
                     &\ + \frac{(-G^{\text{SS}}_{k, 1 3} + i G^{\text{SS}}_{k, 1 4}) (G^{\text{SS}}_{k, 3 1} + i G^{\text{SS}}_{k, 4 1})}{G^{\text{SS}}_{k, 3 3} - i G^{\text{SS}}_{k, 3 4} + i G^{\text{SS}}_{k, 4 3} + G^{\text{SS}}_{k, 4 4}} \\
                     &\ + \frac{(-G^{\text{SS}}_{k, 2 3} + i G^{\text{SS}}_{k, 2 4}) (G^{\text{SS}}_{k, 3 2} + i G^{\text{SS}}_{k, 4 2})}{G^{\text{SS}}_{k, 3 3} - i G^{\text{SS}}_{k, 3 4} + i G^{\text{SS}}_{k, 4 3} + G^{\text{SS}}_{k, 4 4}}. \label{equation:translation_formula}
  \end{split}
\end{equation}
The explicit form of the above translation formula \eqref{equation:translation_formula} and the components of Green's function depend on the choice of the spinor representation.
However, their representation dependencies cancel each other, so the trace of Green's function is invariant under a change of spinor representation.
The method in this subsection can be applied to other quantizations.

\section{Trial holographic models for the Kondo lattice}
\label{appendix:trial_holographic_models_for_the_kondo_lattice}

\begin{figure}[t]
  \centering
  \begin{subfigure}{1.9 in}
    \centering
    \includegraphics[width = 1.9 in]{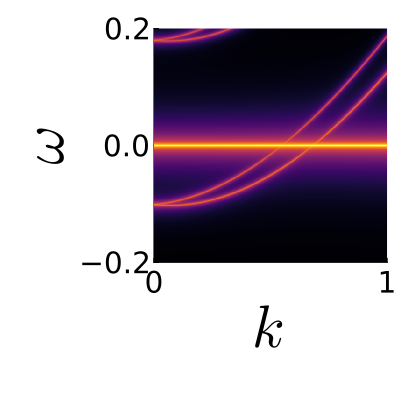}
    \caption{\small $A(\omega, k), V = 0, u_{h} = 10$.}
  \end{subfigure}
  \begin{subfigure}{1.9 in}
    \centering
    \includegraphics[width = 1.9 in]{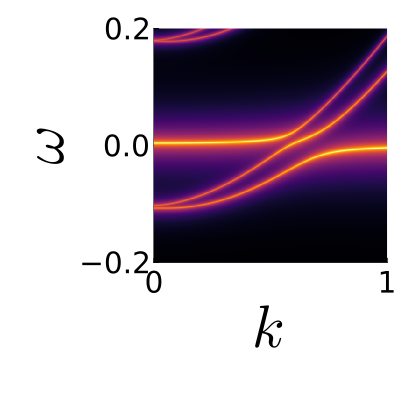}
    \caption{\small $A(\omega, k), V = 0.5, u_{h} = 10$.}
  \end{subfigure}
  \caption{\small
    Spectral function with scalar type interaction.
    The parameters except $g_{1} = 8.2$ are same with figure \ref{figure:holographic_kondo_lattice_model}.
  }
  \label{figure:spectral_function_with_scalar_type_gap}
\end{figure}

\begin{figure}[t]
  \centering
  \begin{subfigure}{1.9 in}
    \centering
    \includegraphics[width = 1.9 in]{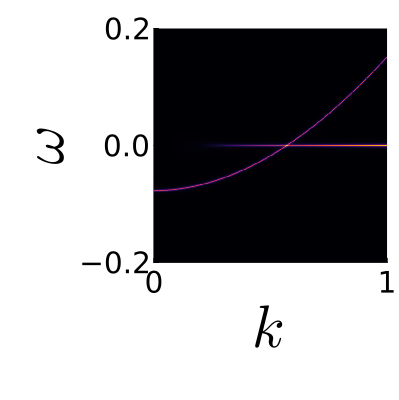}
    \caption{\small $A(\omega, k), V = 0, u_{h} = 10$.}
  \end{subfigure}
  \begin{subfigure}{1.9 in}
    \centering
    \includegraphics[width = 1.9 in]{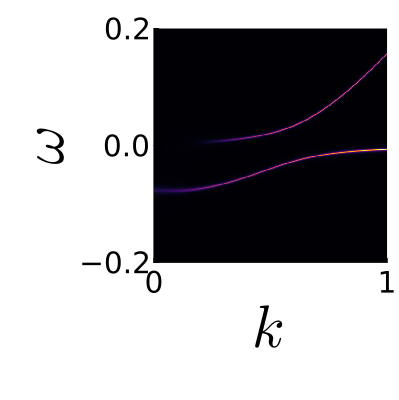}
    \caption{\small $A(\omega, k), V = 0.5, u_{h} = 10$.}
  \end{subfigure}
  \caption{\small
    Spectral function without band isolation.
    The parameters except $g_{2} = 0$ are same with figure \ref{figure:holographic_kondo_lattice_model}.
  }
  \label{figure:spectral_function_without_purifying}
\end{figure}

\subsection{With $g_{1} \Phi_{\text{ps}} \cdot i \mathbb{I}_{4}$ rather than $g_{1} \Phi_{\text{ps}} \cdot \Gamma^{5}$}

When we set
\begin{equation}
  \begin{split}
    S_{\text{spin}} = &\ S_{\text{spin,bdy}} + \sum_{j = 1}^{2} \int \dd^{4} x \sqrt{-g} i \bar{\psi}^{(j)} \left[ \frac{1}{2} \left( \overrightarrow{\slashed{D}}^{(j)} - \overleftarrow{\slashed{D}}^{(j)} \right) - m_{j} \right] \psi^{(j)} \\
                      &\ + \int \dd^{4} x \sqrt{-g} \mqty(\bar{\psi}^{(1)} \\ \bar{\psi}^{(2)})^{T} \mqty(g_{1} \Phi_{\text{ps}} \cdot i \mathbb{I}_{4} & V \Phi_{\text{s}} \cdot i \mathbb{I}_{4} \\ V \Phi_{\text{s}} \cdot i \mathbb{I}_{4} & g_{2} \Phi_{\text{s}} \cdot i \mathbb{I}_{4}) \mqty(\psi^{(1)} \\ \psi^{(2)}),
  \end{split}
\end{equation}
instead of eq. \eqref{equation:action_spinor_term}, eqs. \eqref{equation:differential_equation_matrix_1}, \eqref{equation:differential_equation_matrix_2}, \eqref{equation:differential_equation_matrix_3}, and \eqref{equation:differential_equation_matrix_4} are replaced by
\begin{align}
    \tilde{\mathcal{D}}_{k}^{1} = &\ \mqty(-\frac{(\omega + q_{1} A_{t}) \sigma_{0}}{f \sqrt{\chi}} - \frac{k_{x} \sigma_{3} + k_{y} \sigma_{1}}{\sqrt{f}} & 0 \\ 0 & -\frac{(\omega + q_{2} A_{t}) \sigma_{0}}{f \sqrt{\chi}} - \frac{k_{x} \sigma_{3} + k_{y} \sigma_{1}}{\sqrt{f}}), \\
  \tilde{\mathcal{D}}_{k}^{2} = &\ \mqty(-\frac{(m_{1} - g_{1} \Phi_{\text{ps}}) \sigma_{0}}{u \sqrt{f}} & \frac{V \Phi_{\text{s}} \sigma_{0}}{u \sqrt{f}} \\ \frac{V \Phi_{\text{s}} \sigma_{0}}{u \sqrt{f}} & -\frac{(m_{2} - g_{2} \Phi_{\text{s}}) \sigma_{0}}{u \sqrt{f}}), \\
  \tilde{\mathcal{D}}_{k}^{3} = &\ \mqty(\frac{(m_{1} - g_{1} \Phi_{\text{ps}}) \sigma_{0}}{u \sqrt{f}} & -\frac{V \Phi_{\text{s}} \sigma_{0}}{u \sqrt{f}} \\ -\frac{V \Phi_{\text{s}} \sigma_{0}}{u \sqrt{f}} & \frac{(m_{2} - g_{2} \Phi_{\text{s}}) \sigma_{0}}{u \sqrt{f}}), \\
  \tilde{\mathcal{D}}_{k}^{4} = &\ \mqty(\frac{(\omega + q_{1} A_{t}) \sigma_{0}}{f \sqrt{\chi}} - \frac{k_{x} \sigma_{3} + k_{y} \sigma_{1}}{\sqrt{f}} & 0 \\ 0 & \frac{(\omega + q_{2} A_{t}) \sigma_{0}}{f \sqrt{\chi}} - \frac{k_{x} \sigma_{3} + k_{y} \sigma_{1}}{\sqrt{f}}),
\end{align}
and the hybridization does not open gap completely (see figure \ref{figure:spectral_function_with_scalar_type_gap}).

\subsection{Without $g_{2} \Phi_{\text{s}} \cdot i \mathbb{I}_{4}$}

When we set
\begin{equation}
  \begin{split}
    S_{\text{spin}} = &\ S_{\text{spin,bdy}} + \sum_{j = 1}^{2} \int \dd^{4} x \sqrt{-g} i \bar{\psi}^{(j)} \left[ \frac{1}{2} \left( \overrightarrow{\slashed{D}}^{(j)} - \overleftarrow{\slashed{D}}^{(j)} \right) - m_{j} \right] \psi^{(j)} \\
                      &\ + \int \dd^{4} x \sqrt{-g} \mqty(\bar{\psi}^{(1)} \\ \bar{\psi}^{(2)})^{T} \mqty(g_{1} \Phi_{\text{ps}} \cdot \Gamma^{5} & V \Phi_{\text{s}} \cdot i \mathbb{I}_{4} \\ V \Phi_{\text{s}} \cdot i \mathbb{I}_{4} & 0) \mqty(\psi^{(1)} \\ \psi^{(2)}),
  \end{split}
\end{equation}
instead of eq. \eqref{equation:action_spinor_term}, the flat band is dim near $\vec{k} = 0$ (see figure \ref{figure:spectral_function_without_purifying}).

\bibliographystyle{jhep}
\bibliography{ref.bib}

\end{document}